\documentclass[aip,rsi,12pt,noshowpacs,unsortedaddress,longbibliography,amsmath,amssymb,a4paper]{revtex4-1}

\usepackage{graphicx}
\usepackage{color}
\usepackage{epsfig}
\usepackage{epstopdf}
\usepackage{natbib}
\usepackage{amsmath}

\draft 

\begin{document}
\title{Unidirectional spin Hall magnetoresistance in ferromagnet/normal metal bilayers}

\author{
\bigskip
\bigskip
Can Onur Avci}
\author{Kevin Garello}
\author{Abhijit Ghosh}
\author{Mihai Gabureac}
\author{Santos F. Alvarado}
\author{Pietro Gambardella}
\affiliation{\bigskip
Department of Materials, ETH Z\"{u}rich, H\"{o}nggerbergring 64, CH-8093 Z\"{u}rich, Switzerland}


\begin{abstract}
\bigskip
\bigskip
\textbf{Magnetoresistive effects are usually invariant upon inversion of the magnetization direction. In noncentrosymmetric conductors, however, nonlinear resistive terms can give rise to a current dependence that is quadratic in the applied voltage and linear in the magnetization. Here we demonstrate that such conditions are realized in simple bilayer metal films where the spin-orbit interaction and spin-dependent scattering couple the current-induced spin accumulation to the electrical conductivity. We show that the longitudinal resistance of Ta$|$Co and Pt$|$Co bilayers changes when reversing the polarity of the current or the sign of the magnetization. This unidirectional magnetoresistance scales linearly with current density and has opposite sign in Ta and Pt, which we associate to the modification of the interface scattering potential induced by the spin Hall effect in these materials. Our results suggest a new route to detect magnetization switching in spintronic devices using a two-terminal geometry, which applies also to heterostructures including topological insulators.}
\end{abstract}

\maketitle 

The effects of the magnetization on the electric conductivity of metals have been studied for a long time~\cite{Thomson1856prsl}, providing understanding of fundamental phenomena associated to electron transport and magnetism as well as manyfold applications in sensor technology. The anisotropic magnetoresistance (AMR), the change of the resistance of a material upon rotation of the magnetization, is a prominent manifestation of spin-orbit coupling and spin-dependent conductivity in bulk ferromagnets~\cite{Campbell1970jpc,Mcguire1975ieeetm}. In thin film heterostructures, the additional possibility of orienting the magnetization of stacked ferromagnetic layers parallel or antiparallel to each other
gives rise to the celebrated giant magnetoresistance (GMR) effect~\cite{BaibichPRL1988Gmr,Grunberg89prb}, which has played a major role in all modern developments of spintronics. Together with the early spin injection experiments~\cite{JohnsonPRL1985,Jedema2001N}, the study of GMR revealed how nonequilibrium spin accumulation at the interface between ferromagnetic (FM) and normal metal (NM) conductors governs the propagation of spin currents~\cite{Johnson1987PRB,vanSon1987PRL,Valet1993PRB,maekawa2012book} and, ultimately, the conductivity of multilayer systems~\cite{Valet1993PRB,Brataas2006PR}.

Recently, it has been shown that significant spin accumulation at a FM$|$NM interface can be achieved using a current-in-plane (CIP) geometry owing to the spin Hall effect (SHE) in the NM~\cite{Sinova2014Arxiv}. When FM is a metal and NM is a heavy element such as Pt or Ta, the spin accumulation is strong enough to induce magnetization reversal of nm-thick FM layers at current densities of the order of $j=10^7-10^8$~A/cm$^2$ (Refs.~\onlinecite{MironN2011,LiuS2012,GarelloNN2013}). When FM is an insulator, such as yttrium iron garnet,
the SHE causes an unusual magnetoresistance associated to the back-flow of a spin current into the NM when the spin accumulation $\boldsymbol{\mu}_s \sim (\mathbf{j}\times \hat{\mathbf{z}})$ is aligned with the magnetization of the FM, which increases the conductivity of the NM due to the inverse SHE~\cite{NakayamaPRL2013,HahnPRB2013,althammer2013PRB,Miao2014PRL}. This so-called spin Hall magnetoresistance (SMR) is characterized by $R^y < R^z \approx R^x$, where $R^i$ is the resistance measured when the magnetization ($\mathbf{M}$) is saturated parallel to $i=x,y,z$, and differs from the conventional AMR in polycrystalline samples, for which $R = R^x - (R^{y,z}-R^x)[\hat{\mathbf{M}}\cdot\hat{\mathbf{j}}]^2$ and $R^y \approx R^z < R^x$ (Ref.~\onlinecite{Mcguire1975ieeetm}).

\begin{figure}
  \centering
  \includegraphics[width=14 cm]{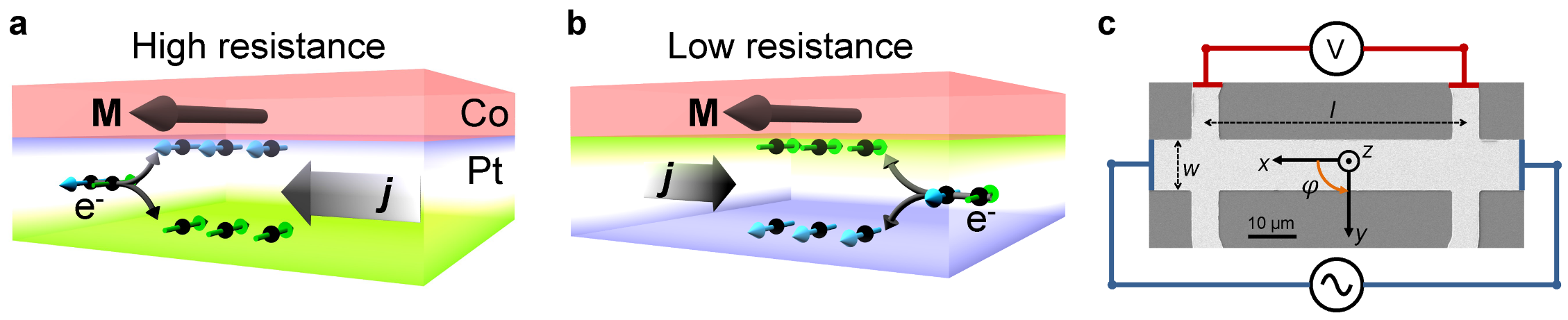}\\
  \caption{\textbf{Illustration of the unidirectional spin Hall magnetoresistance effect and sample layout.} \textbf{a,} Parallel alignment of the SHE-induced nonequilibrium magnetization at the FM$|$NM interface with the magnetization of the FM increases the resistivity of the bilayer. \textbf{b,} Antiparallel alignment decreases the resistivity. The arrows indicate the direction of the spin magnetic moment. \textbf{c,} Scanning electron micrograph of a sample and schematics of the longitudinal resistance measurements.}\label{fig1}
\end{figure}

In this work, we report on a new magnetoresistance effect occurring in FM$|$NM bilayers with the NM possessing large SHE. The effect combines features that are typical of the current-in-plane (CIP) GMR and SHE, whereby the spin accumulation induced by the SHE in the NM replaces one of the FM polarizers of a typical GMR device. Differently from GMR, however, this effect introduces a nonlinear dependence of the resistance on the current, which gives it unique unidirectional properties: the resistivity changes when reversing either the sign of the magnetization or the polarity of the current, increasing (decreasing) when the SHE-induced non equilibrium magnetization at the FM$|$NM interface is oriented parallel (antiparallel) to the magnetization of the FM, as illustrated in Fig.~\ref{fig1}a,b. We associate this phenomenon to the modulation of the FM$|$NM interface resistance due to the SHE-induced spin accumulation, which gives rise to a nonlinear contribution to the longitudinal conductivity that scales proportionally with the current density and has opposite sign in Pt and Ta. Contrary to the linear magnetoresistive effects, including the AMR, GMR, and SMR described above, which are even with respect to the inversion of either the current or magnetization owing to the time reversal symmetry embodied in the Onsager's reciprocity relations, here we observe $R(\mathbf{j},\mathbf{M})=-R(\mathbf{-j},\mathbf{M})=-R(\mathbf{j},\mathbf{-M})$, providing a unidirectional contribution to the magnetoresistance in simple bilayer systems.

\begin{figure}
  \centering
  \includegraphics[width=16.5 cm]{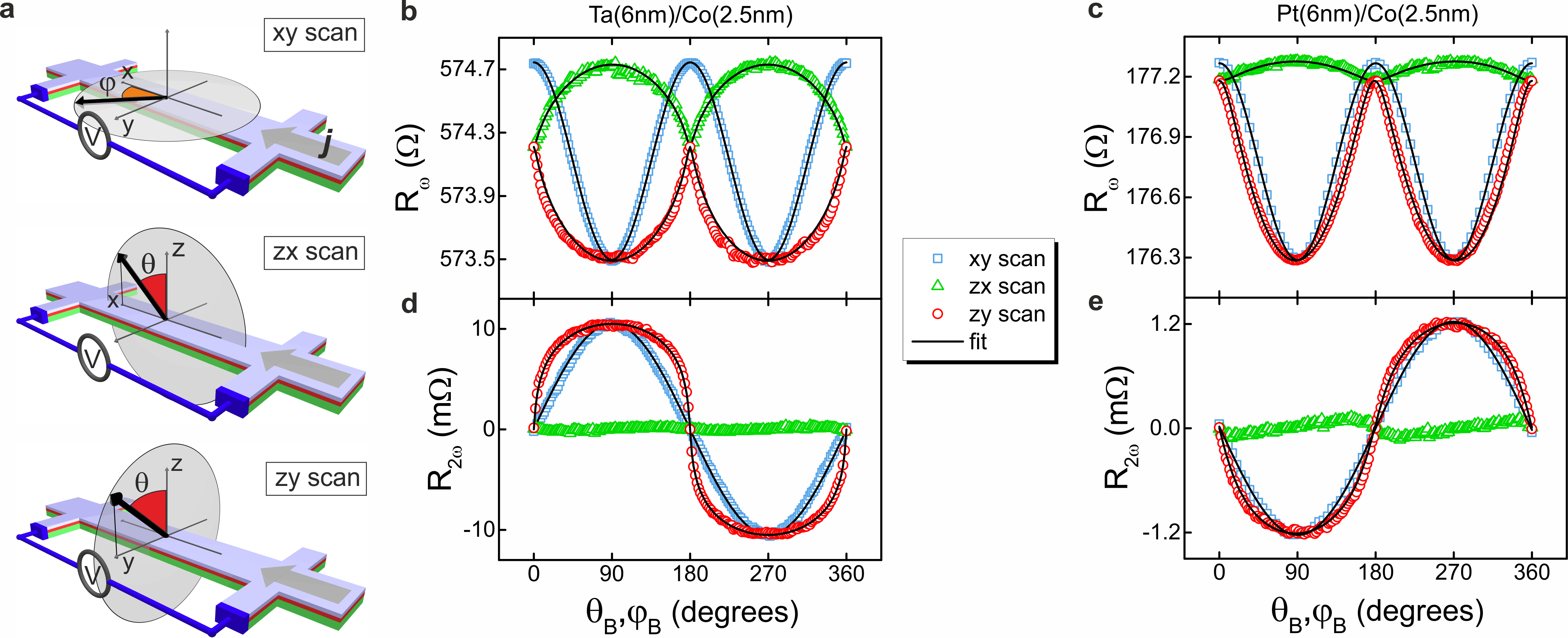}\\
  \caption{\textbf{Linear and nonlinear magnetoresistance.} \textbf{a,} Geometry of the measurements. \textbf{b,} First harmonic resistance of Ta(6nm)$|$Co(2.5nm) and \textbf{c,} Pt(6nm)$|$Co(2.5nm) measured with a current density of $j=10^7$~A/cm$^2$. \textbf{d,e} Second harmonic resistance measured simultaneously with \textbf{b,c}. The dimensions of the Hall bars are $l=50$~$\mu$m and $w=10$~$\mu$m.}
  \label{fig2}
\end{figure}

\bigskip \noindent \textbf{Sample layout}\\
The samples studied in this work are Pt(1-9nm)$|$Co(2.5nm) and Ta(1-9nm)$|$Co(2.5nm) films with spontaneous in-plane magnetization, capped by 2~nm AlO$_x$ and patterned in the shape of Hall bars of nominal length $l=20-50$~$\mu$m, width $w=4-10$~$\mu$m, and $l/w = 4$, as shown in Fig.~\ref{fig1}c.
Additional control experiments were carried out on single Co, Ta, and Pt films, and Ta(1,6nm)$|$Cu(2,4,6nm)$|$Co(2.5nm) trilayers. To characterize the magnetic and electrical properties of these layers we performed harmonic measurements of the longitudinal resistance ($R$, see Supplementary Information) and transverse Hall resistance ($R_H$)~\cite{GarelloNN2013,AvciPRB2014,HayashiPRB2014,AvciPRB2014b} as a function of a vector magnetic field defined by the polar and azimuthal coordinates $\theta_{B}$ and $\varphi_{B}$. The measurements were carried out at room temperature by injecting an ac current of frequency $\omega / 2\pi=10$~Hz and simultaneously recording the first ($R_{\omega}$) and second harmonic resistance ($R_{2\omega}$) between the contacts shown in Fig.~\ref{fig1}c while rotating the sample in a uniform magnetic field of 1.7~T. Here, $R_{\omega}$ represents the linear response of the samples to the current, that is, the conventional resistance. In order to include the different magnetoresistive angular dependencies in a single expression we write this term as
\begin{equation}
\label{f_MR}
R_{\omega}=R^{z}+(R^{x}-R^{z})\sin^{2}\theta\cos^{2}\varphi+(R^{y}-R^{z})\sin^{2}\theta\sin^{2}\varphi \, ,
\end{equation}
where $\theta$ and $\varphi$ are the polar and azimuthal angles of $\mathbf{M}$, as schematized in Fig.~\ref{fig2}a.
$R_{2\omega}$ describes resistance contributions that vary quadratically with the applied voltage and includes the current-induced changes of resistivity that are the main focus of this work.

\bigskip \noindent \textbf{Magnetoresistance measurements}\\
Figure~\ref{fig2}b and c show the resistance of Ta(6nm)$|$Co(2.5nm) and Pt(6nm)$|$Co(2.5nm) layers during rotation of the applied field in the $xy$, $zx$, and $zy$ planes. We observe a sizeable magnetoresistance (MR) in all three orthogonal planes and $R^x > R^z > R^y$ for both samples, in agreement with previous measurements on Pt$|$Co films\cite{KobsPRL2011,LuPRB2013}. The resistivity of Ta(6nm)$|$Co(2.5nm) [Pt(6nm)$|$Co(2.5nm)] is 108.9 (36.8)~$\mu\Omega$cm and the MR ratios are $(R^x-R^z)/R^z=0.09\%$ [$0.05\%$] and $(R^z-R^y)/R^z=0.12\%$ [$0.53\%$], showing a large SMR-like behavior compared to Ta$|$YIG and Pt$|$YIG\cite{NakayamaPRL2013,HahnPRB2013}.
The solid lines represent fits to the MR using Eq.~\ref{f_MR} and $\theta$ simultaneously measured via the anomalous Hall resistance (see Supplementary Information).

In addition to the linear resistance, we measure an unexpected nonlinear resistance, $R_{2\omega}$, which has a different angular dependence compared to $R_{\omega}$ and opposite sign in Pt and Ta, as shown in Fig.~\ref{fig2}d and e. By fitting the curves with respect to the angles $\theta$ and $\varphi$ (solid lines), we find that $R_{2\omega} \sim sin\theta\sin\varphi \sim M_y$. In the following, we discuss the type of nonlinear effects that can give rise to such a symmetry.

\bigskip \noindent \textbf{Spin-orbit torques and thermoelectric contributions}\\
First, we consider oscillations of the magnetization due to the current-induced spin-orbit torques (SOT)~\cite{GarelloNN2013,AvciPRB2014,HayashiPRB2014,KimNM2013,AvciPRB2014b}. As the SOT are proportional to the current, ac oscillations of the magnetization can introduce second-order contributions to $R_{2\omega}$ due to the first order MR described by Eq.~\ref{f_MR}. However, as shown in detail in the Supplementary Information, the SOT-induced signal is not compatible with the angular dependence of $R_{2\omega}$. Both the field-like and antidamping-like SOT (as well as the torque due to the Oersted field) vanish for $\mathbf{M} \parallel y$, where $|R_{2\omega}|$ is maximum. Moreover, the field-like SOT is small in 2.5~nm thick Co layers~\cite{AvciPRB2014b}, whereas the antidamping SOT can only induce variations of $R_{2\omega}$ in the $zx$ plane with maxima and minima close to $\theta_B = 0^{\circ}$ and $180^{\circ}$, which we observe to be small and more pronounced in Pt$|$Co relative to Ta$|$Co (Fig.~\ref{fig2}d and e bottom panels).

Second, we analyze the influence of thermal gradients ($\nabla T$) and related thermoelectric effects. The anomalous Nernst (ANE) and spin Seebeck effect (SSE)~\cite{WeilerPRL2012,KikkawaPRL2013}, both inducing a longitudinal voltage proportional to $j^2 (\mathbf{M }\times \nabla T)$, can give rise to a similar angular dependence as observed for $R_{2\omega}$ when $\nabla T \parallel \hat{\mathbf{z}}$ (see Supplementary Information). Here, we find that thermoelectric voltages are negligible in Pt$|$Co, in agreement with the very small thermal gradients reported for this system\cite{AvciPRB2014b}. In Ta$|$Co, on the other hand, the much larger resistivity of Ta relative to Co results in a higher current flowing in the Co layer and a positive $\nabla T$. In such a case, the second harmonic signal of thermal origin, $R^{\nabla T}_{2\omega}$, can be simply estimated from its transverse (Hall) counterpart scaled by the geometric factor $l/w$ when the magnetization is tilted in the $x$ direction, and subtracted from the raw $R_{2\omega}$ signal. Accordingly, we find that $R^{\nabla T}_{2\omega} = 5$~m$\Omega$ in Ta(6nm)$|$Co(2.5nm), which accounts for only about $50\%$ of the total $R_{2\omega}$ reported in Fig.~\ref{fig2}. The same procedure applied to Pt$|$Co gives $R^{\nabla T}_{2\omega}$~m$\Omega$ of the order of 5\% of the total $R_{2\omega}$, whereas in the control samples lacking a heavy metal we find uniquely a signal of thermal (ANE) origin. We conclude that there is an additional magnetoresistive effect in the Pt$|$Co and Ta$|$Co bilayers that cannot be accounted for by either current-induced magnetization dynamics or thermoelectric voltages.

\bigskip \noindent \textbf{Unidirectional spin Hall magnetoresistance}\\
The symmetry as well as the opposite sign of the nonlinear resistance in Ta$|$Co and Pt$|$Co suggest that it is related to the scalar product of the magnetization with the SHE-induced spin accumulation at the FM$|$NM interface, $(\mathbf{j}\times \hat{\mathbf{z}}) \cdot \mathbf{M}$, giving rise to a chiral MR contribution $R^{USMR}_{2\omega} \sim \mathbf{j}\times \mathbf{M}$. This relation describes the general features expected from a unidirectional magnetoresistance driven by the spin Hall effect (USMR). We note that this MR contribution depends on the current direction and that the resistance of the bilayer increases when the direction of the majority spins in the FM and the spin accumulation vector are parallel to each other, and decreases when they are antiparallel. This may appear counterintuitive at first sight, considering that the conductivity of Co is larger for the majority spins. However, as we will discuss later, this behavior is consistent with the theory of GMR in FM$|$NM$|$FM heterostructures~\cite{Valet1993PRB,CamleyPRL1989,HoodPRB1992} when only a single FM$|$NM interface is retained and the SHE is taken into account.

\begin{figure}
  \centering
  \includegraphics[width=12 cm]{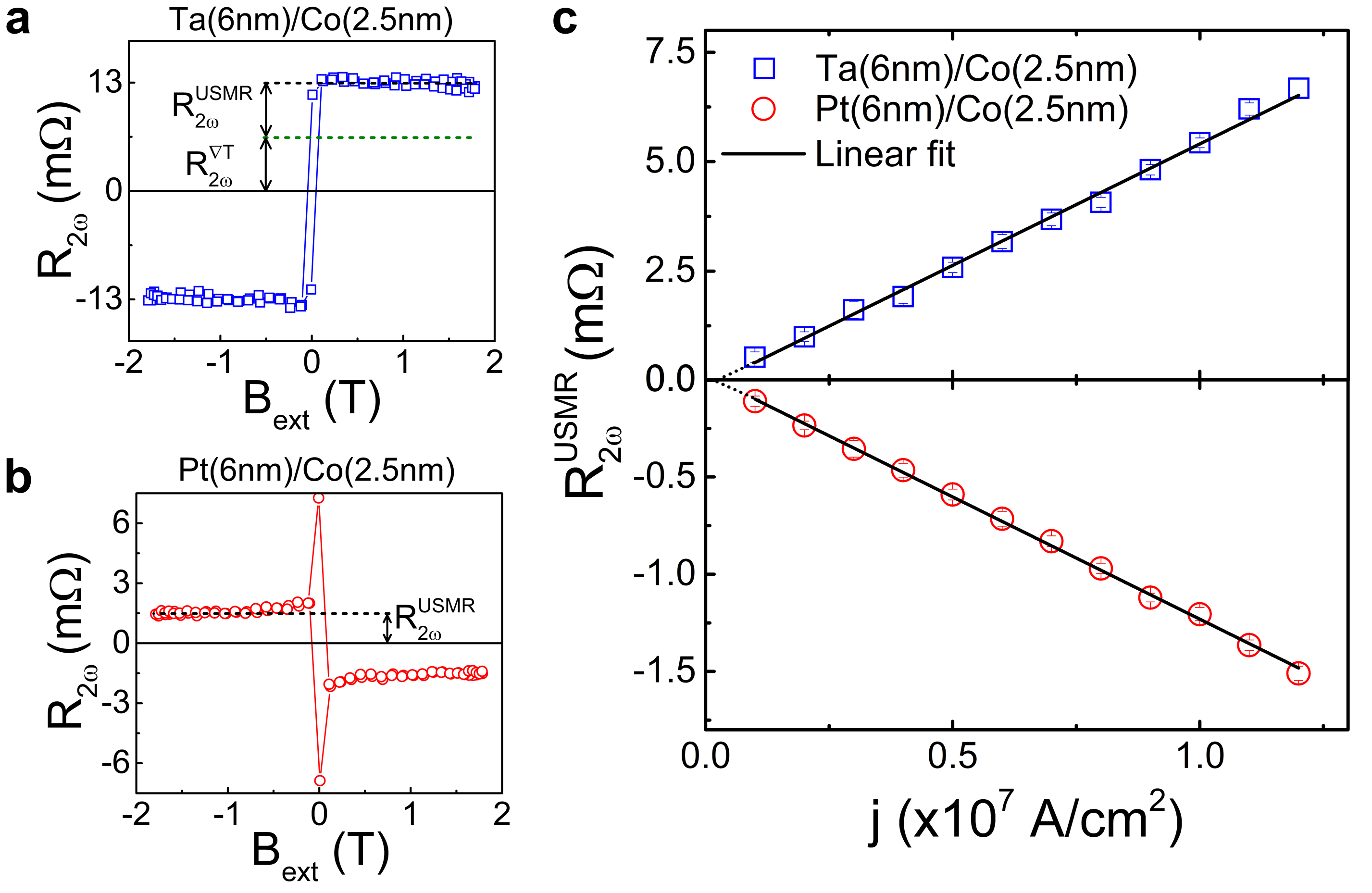}\\
  \caption{\textbf{Field and current dependence of the nonlinear magnetoresistance.} $R_{2\omega}$ of \textbf{a,} Ta(6nm)$|$Co(2.5nm) and \textbf{b,} Pt(6nm)$|$Co(2.5nm) recorded during a field sweep along $y$ with a current density of $j=1.2\cdot10^{7}$~A/cm$^2$. \textbf{c,} Current dependence of $R^{USMR}_{2\omega}$. The solid lines are fits to the data. The slope gives the amplitude of the USMR, which is 5.5~m$\Omega$ and 1.25~m$\Omega$ per $10^7$~A/cm$^2$ for these samples. The thermal contribution $R^{\nabla T}_{2\omega}$ has been subtracted from the Ta$|$Co data. The dimensions of the Hall bars are $l=50$~$\mu$m and $w=10$~$\mu$m.}\label{fig3}
\end{figure}

To investigate further the USMR we have measured $R_{2\omega}$ as a function of an external magnetic field applied parallel to $\hat{\mathbf{y}}$ and current amplitude. Figure~\ref{fig3}a shows that $R_{2\omega}$ is constant as a a function of field for Ta$|$Co as well as for Pt$|$Co (Fig.~\ref{fig3}b) and reverses sign upons switching of the magnetization from the $y$ to the $-y$ direction. In the Pt$|$Co case we observe also two spikes, which we attribute to the magnetization breaking into domains at low field and giving rise to dynamic effects on the domain walls~\cite{GarelloNN2013}. Note that the magnetization of Pt$|$Co is not fully saturated below 0.65~T due to the large perpendicular magnetic anisotropy of this system, differently from Ta$|$Co (Supplementary Information). Figure~\ref{fig3}c shows the current dependence of $R^{USMR}_{2\omega} = R_{2\omega}- R^{\nabla T}_{2\omega}$ ($R^{USMR}_{2\omega} \approx R_{2\omega}$ for Pt$|$Co) obtained by taking the average of the data measured at fields larger than $|\pm1|$~T. $R^{USMR}_{2\omega}$ is linear with the injected current density and converges to zero within the error bar of the linear fit (black lines).

\begin{figure}
  \centering
  \includegraphics[width=12 cm]{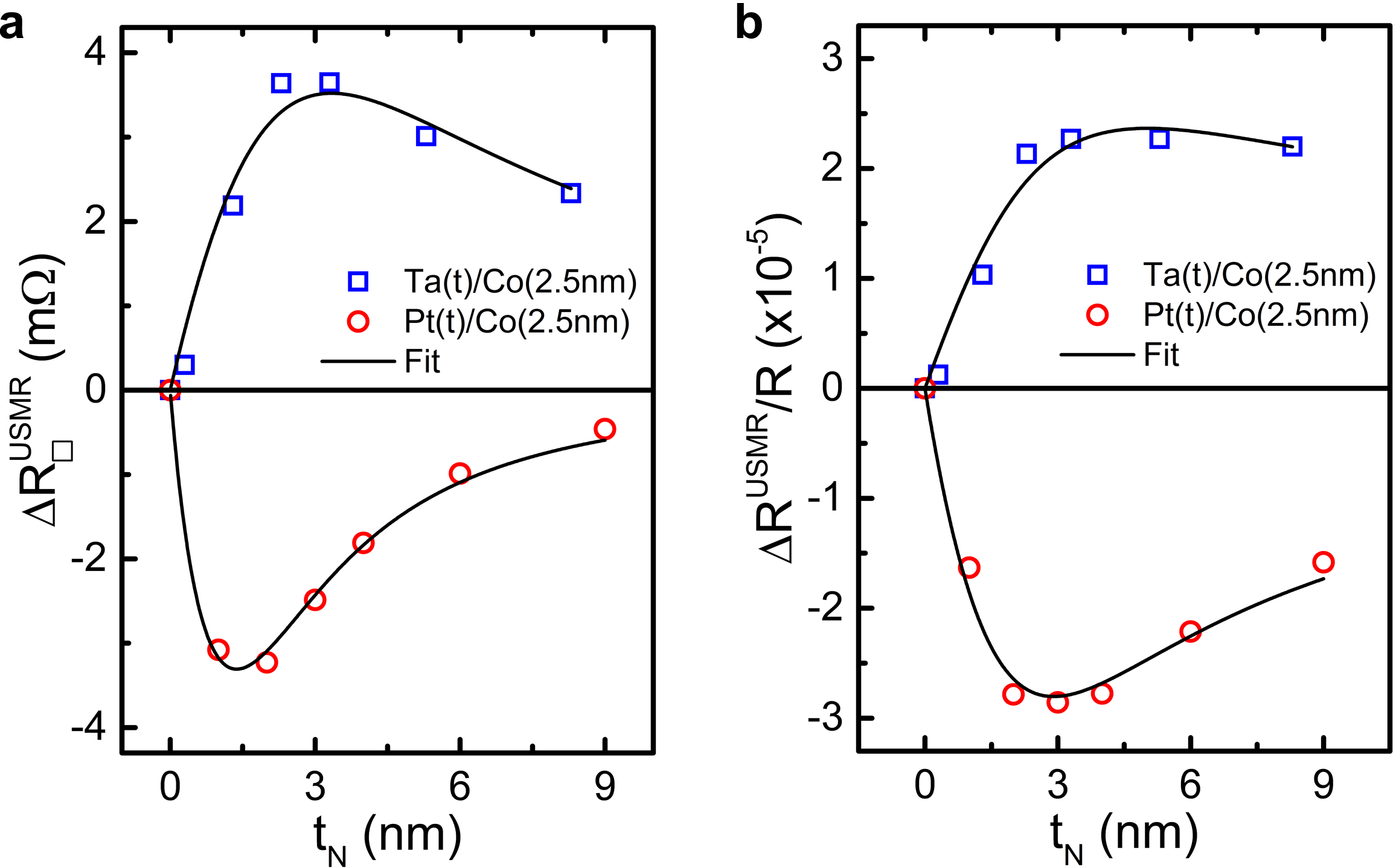}\\
  \caption{\textbf{USMR as a function of NM thickness.} \textbf{a,} Sheet resistance $\Delta R^{USMR}_{2\omega}$ as a function of Ta (squares) and Pt (circles) thickness measured at constant current density $j=10^7$~A/cm$^2$. The Co layer is 2.5~nm thick in all samples. \textbf{b,} Normalized resistance $\Delta R^{USMR}_{2\omega}/R$. The solid lines are fits to the data according to the model described in the text.}\label{fig4}
\end{figure}

To verify the role of the interfacial spin accumulation due to the SHE
we examined the dependence of the USMR on the thickness of the NM. Figure~\ref{fig4}a and b show the absolute change of sheet resistance $\Delta R^{USMR} = R^{USMR}_{2\omega}(\pm \mathbf{M},\pm \mathbf{j})-R^{USMR}_{2\omega}(\pm \mathbf{M},\mp \mathbf{j})$ and the relative change of resistivity $\Delta R^{USMR}/R$ measured at constant current density as a function of the Ta and Pt thickness. Both curves exhibit qualitatively similar behavior: an initial sharp increase below 2-3~nm and a gradual decrease as the NM layer becomes thicker. We note that the USMR signal is almost absent in Ta(1nm)$|$Co, contrary to Pt(1nm)$|$Co, which we attribute to the oxidation of Ta when deposited on SiO$_2$ and its consequent poor contribution to electrical conduction. The initial increase of the USMR is consistent with the increment of the spin accumulation at the FM$|$NM interface as the thickness of the NM becomes larger than the spin diffusion length, which is of the order of 1.5~nm in both Ta and Pt~\cite{HahnPRB2013,ZhangAPL2013}. Moreover, we observe that the decline of the signal in the thicker samples is stronger in Pt$|$Co than in Ta$|$Co. This behavior is consistent with Pt gradually dominating the conduction due to its low resistivity, and a smaller proportion of the current experiencing interface scattering in Pt$|$Co. Conversely, the high resistivity of Ta shunts the current towards the Co side, increasing the relative proportion of the current affected by scattering at the Ta$|$Co interface.

As an additional check to validate these arguments we have performed measurements on single Ta(6nm), Pt(6nm), and Co(8nm) layers as well as on Ta(1,6nm)$|$Cu(2,4,6nm)$|$Co(2.5nm) trilayers, all capped by 2~nm AlO$_x$. The USMR is absent in the Ta, Pt, and Co single layers, which also excludes any self induced magnetoresistive effect~\cite{DyakonovPRL2007} and proves the essential role of the FM$|$NM interface. On the other hand, we find a sizable USMR when a Cu spacer is inserted between Ta and Co, which excludes proximity-induced magnetization as a possible cause for the USMR (see Supplementary Information).

\begin{figure}
  \centering
  \includegraphics[width=14 cm]{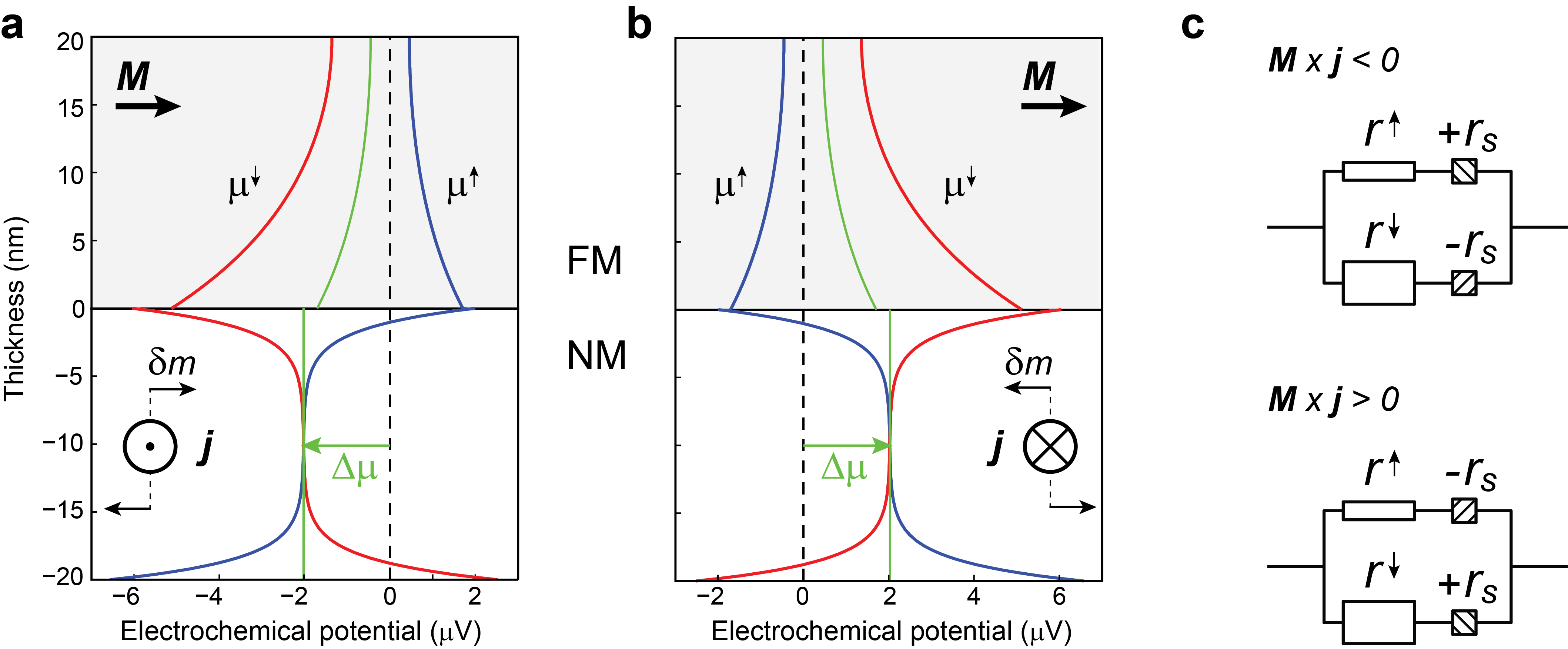}\\
  \caption{\textbf{Modulation of the spin accumulation, spin-dependent electrochemical potential, and interface resistance by the SHE.} Profile of the electrochemical potential of majority ($\mu^\uparrow$, blue lines) and minority ($\mu^\downarrow$, red lines) electrons in proximity of the FM$|$NM interface for \textbf{a,} positive and \textbf{b,} negative current. The electrochemical potential of the NM shifts relative to that of the FM as indicated by the green arrow. The direction of the magnetization is $\mathbf{M} \parallel \hat{\mathbf{y}}$ in both panels. In our notation, the majority spins are oriented antiparallel to $\mathbf{M}$ and, likewise, the non-equilibrium magnetization induced by the SHE has opposite sign relative to $\boldsymbol{\mu}_s$. Reversing $\mathbf{M}$ is equivalent to exchanging $\mu^\uparrow$ and $\mu^{\downarrow}$ and inverting $\Delta \mu$. The sign of the SHE corresponds to Pt$|$Co; the parameters used to calculate $\mu^\uparrow$ and $\mu^\downarrow$ are given in the Supplementary Information. \textbf{c,} Two current series resistor model of the USMR corresponding to panel \textbf{a} (top, higher resistance) and \textbf{b} (bottom, lower resistance). Note that the resistances $r^\uparrow$ and $r^\downarrow$ may be generalized to include also the bulk spin-dependent resistances of the FM layer.}\label{fig5}
\end{figure}

\bigskip \noindent \textbf{Discussion}\\
Based on the analysis presented above, we conclude that the current-induced spin accumulation creates an additional spin-dependent interface resistance that adds or subtracts to the total resistance depending on sign of the cross product $\mathbf{j}\times \mathbf{M}$. Given the in-plane geometry, the interpretation of this effect requires a Boltzmann equation approach to model the spin- and momentum-dependent reflection and transmission of electrons at the FM$|$NM interface, equivalent to extending the theory of CIP-GMR~\cite{CamleyPRL1989,HoodPRB1992} beyond first order and including the SHE. However, a qualitative understanding of the USMR can be derived also by introducing a two current series resistor model and an interface resistance term proportional to the SHE-induced shift of the electrochemical potential between the FM and NM layers. The latter can be calculated using a one-dimensional drift-diffusion approach\cite{vanSon1987PRL,Valet1993PRB,NakayamaPRL2013}. We consider two separate conduction channels for the minority (spin $\uparrow$) and majority (spin $\downarrow$) electrons. As in bulk FM, scattering at the interface is spin-dependent due to the unequal density of majority and minority states near the Fermi level, which, in most cases, leads to a larger resistance for minority electrons relative to majority electrons: $r^\downarrow > r^\uparrow$. This resistance difference is at the heart of the GMR effect, both in the CIP\cite{CamleyPRL1989,HoodPRB1992,dieny92jpcm} and the current-perpendicular-to-plane (CPP) geometry~\cite{Valet1993PRB,bass2014jmmm}. Additionally, when an electric current flows from a FM to a NM or viceversa, another resistance term appears due to the conductivity mismatch between majority and minority electrons on opposite sides of the junction, which results in spin accumulation (Refs.~\onlinecite{Johnson1987PRB,vanSon1987PRL}). This so-called "spin-coupled interface resistance", plays a role in CPP-GMR as well as in local and nonlocal spin injection devices\cite{Valet1993PRB,Jedema2001N,Valenzuela2006N}, whereas in the CIP geometry it is normally neglected because there is no net charge flow across the interface and the spin accumulation is assumed to be zero. If we take the SHE into account, however, the transverse spin current flowing between the NM and the FM induces a splitting of the spin-dependent electrochemical potentials $\mu^\uparrow$ and $\mu^\downarrow$ and a net interfacial spin accumulation $\mu_s = \mu^{\uparrow}-\mu^{\downarrow}$, which is given by
\begin{equation}
      \mu_{s_N}=\mu_{s_N}^0\tanh\frac{t_N}{2\lambda_N}\frac{1+\frac{r_b}{\rho_F\lambda_F}(1-P^2)\tanh\frac{t_F}{\lambda_F}}
      {1+\left(\frac{\rho_N\lambda_N}{\rho_F\lambda_F}\coth\frac{t_n}{\lambda_N} - \frac{r_b}{\rho_F\lambda_F}\right)(1-P^2)\tanh\frac{t_F}{\lambda_F}},
      \label{muN0}
\end{equation}
where $\mu_{s_N}^0 = 2\,e \,\rho_N\lambda_N \,\theta_{SH}\,j$ is the bare spin accumulation due to the SHE that would occur in a single, infinitely thick NM layer, $\theta_{\scriptscriptstyle{SH}}$ the spin Hall angle of the NM, $\rho_{N,F}$ and $\lambda_{N,F}$ are the resistivity and spin diffusion length of the NM and FM, respectively, and $r_b = (r^\uparrow + r^\downarrow)/4$ is the interface resistance\cite{Valet1993PRB}. Moreover, the same effect induces a shift $\Delta \mu = \mu_N - \mu_F$ of the electrochemical potential $\mu_{N,F} = (\mu_{N,F}^{+}+\mu_{N,F}^{-})/2$ of the NM relative to the FM:
\begin{equation}
      \Delta\mu_N  =-(P+\gamma \tilde{r})\mu_{s_N}^0\tanh\frac{t_N}{2\lambda_N}\frac{1}{1+\tilde{r}}\frac{1}
      {1+\frac{\frac{\rho_N\lambda_N}{\rho_F\lambda_F}(1-P^2)\tanh\frac{t_F}{\lambda_F}\coth\frac{t_n}{\lambda_N}}{1-\tilde{r}}},
      \label{DeltamuN}
\end{equation}
where $\gamma = (r^\downarrow -r^\uparrow)/(r^\uparrow+r^\downarrow)$ and $\tilde{r}=\frac{r_b}{\rho_F\lambda_F}(1-P^2)\tanh\frac{t_F}{\lambda_F}$. Figure~\ref{fig5}a and b show a graphical representation of $\mu_s$ and $\Delta\mu_N$; details about the derivation of Eqs.~\ref{muN0} and~\ref{DeltamuN} are given in the Supplementary Information. A key point is that $\Delta \mu_{N}$ depends on the product $P\theta_{\scriptscriptstyle{SH}}j$, as the USMR, and is linked with the spin-dependent scattering potential that gives rise to different transmission coefficients for majority and minority electrons at the FM$|$NM interface~\cite{zhang1992prb}. We can thus draw the following qualitative interpretation of the USMR: when the nonequilibrium magnetization induced by the SHE and the magnetization of the FM are parallel to each other ($\delta \mathbf{m} \parallel \mathbf{M}$), the transmission of $\uparrow$ ($\downarrow$) electrons across the interface is reduced (enhanced) by the accumulation of majority electrons at the FM$|$NM boundary, due to the conductivity mismatch of $\uparrow$ and $\downarrow$ spins in the two materials. Likewise, when $\delta \mathbf{m} \parallel -\mathbf{M}$, the transmission of $\downarrow$ ($\uparrow$) electrons across the interface is reduced (enhanced) since minority electrons accumulate at the FM$|$NM boundary. The overall effect is a modulation of the interface resistance of the $\uparrow$ and $\downarrow$ spin channels by a nonlinear term $\pm r_s$, as schematized in Fig.~\ref{fig5}c. This two current ($\uparrow$ and $\downarrow$) series resistor model shows that such a modulation leads to a resistance difference between the two configurations $\Delta R^{USMR}_{2\omega} \sim 2 r_s \gamma$.

Accordingly, using realistic values of $r_b$, $\rho_N$, and $\rho_F$ for Ta$|$Co and Pt$|$Co, we fit the dependence of the USMR on current and NM thickness to the following phenomenological expression (see Supplementary Information): $\Delta R^{USMR} = A \tanh\frac{t_N}{2\lambda_N}/(1+R_{FI}/R_N)^2$, where $A$ is a parameter proportional to $P\mu_{s_N}^0$ representing the amplitude of the effect, $R_{FI}$ is the effective resistance of the FM and interface regions, and $R_N = \rho_N l/(w t _N)$ is the resistance of the NM. The denominator accounts for the decreased fraction of electrons that scatter at the interface as the thickness of the NM increases. Similarly, we obtain $\Delta R^{USMR}/R = (A/R) \tanh\frac{t_N}{2\lambda_N}/(1+R_{FI}/R_N)$. As shown in Fig.~\ref{fig4}, these simple expressions fit $\Delta R^{USMR}$ and $\Delta R^{USMR}/R$ remarkably well, providing also values of $\lambda_{Pt}= 1.1$~nm and $\lambda_{Ta}= 1.4$~nm that are in agreement with previous measurements~\cite{HahnPRB2013,ZhangAPL2013}. Our model thus captures the essential features of the USMR, namely its sign, angular dependence, and proportionality to the current. Detailed calculations including realistic scattering parameters within a nonlinearized Boltzmann approach including the SHE~\cite{HaneyPRB2013} should be able to account for quantitative aspects of the USMR in different materials. We stress also that the USMR is not uniquely linked to the SHE but may arise also due to other sources of nonequilibrium spin accumulation, such as the Rashba effect at FM$|$NM interfaces and topological insulators\cite{HaneyPRB2013,ManchonPRB2008,MironNM2010,mahfouzi2012prl} as well as the anomalous Hall effect in FM.

\bigskip \noindent \textbf{Conclusions}\\
The existence of a nonlinear magnetoresistive term proportional to $\mathbf{j}\times\mathbf{M}$ has both fundamental and practical implications. Identifying which symmetries survive the breakdown of the Onsager relationships in the nonlinear regime is central to the understanding of electron transport phenomena, particularly in mesoscopic and magnetic conductors where such effects can also have thermoelectric and magneto-optical counterparts~\cite{buttiker04prl,wyder01prl}. In this respect, the USMR shows that the longitudinal conductivity has an antisymmetric Hall-like component that has so far remained unnoticed. We expect such a component to be a general feature of noncentrosymmetric magnetic systems with strong spin-orbit coupling. We note also that the USMR
differs from the nonlinear MR observed in chiral conductors, such as metal helices~\cite{wyder01prl} and molecular crystals~\cite{pop14nc}, which is proportional to $\mathbf{j}\cdot\mathbf{M}$.

In the field of spintronics, nonlinear interactions between spin and charge are emerging as a tool to detect spin currents\cite{vera2012NP}  and thermoelectric\cite{slachter2010NP} effects, as well as magnetization reversal in dual spin valves~\cite{AzizPRL2009}. Although the USMR is only a small fraction of the total resistance, its relative amplitude is of the same order of magnitude of the spin transresistance measured in nonlocal metal spin valves\cite{Jedema2001N,Valenzuela2006N}, which is a method of choice for the investigation of spin currents. The thermoelectric counterpart of the USMR, related to the spin Nernst effect, may be used to detect heat-induced spin accumulation by modulation of the magnetization rather than an electric current. We note that the electric field created by the USMR is of the order of 2~V/m per 10$^{7}$~A/cm$^{2}$, which is comparable to the ANE\cite{AvciPRB2014b} and three orders of magnitude larger than the typical electric fields due to the SSE\cite{WeilerPRL2012,KikkawaPRL2013}.

In terms of applications, the USMR may be used to add 360$^{\circ}$ directional sensitivity to AMR sensors, which are widely employed for position, magnetic field, and current sensing, and already include built-in modulation circuitry for accurate resistance measurements. Most interestingly, the USMR shows that it is possible to realize two-terminal spintronic devices where switching is performed by SOT\cite{MironN2011,LiuS2012} and reading by a resistance measurement. Such a scheme involves only one FM layer and minimum patterning effort. Finally, we believe that there is substantial room for improving the amplitude of the USMR to levels closer to the AMR, either by material or heterostructure engineering. In particular, the USMR could increase significantly in magnetic topological insulator structures due to the very large spin accumulation and reduced bulk conductance reported for these systems\cite{ralph14Nature,fan2014NM}.

\clearpage
\bigskip \noindent \textbf{Methods}\\
\textbf{Sample preparation.} The Pt(1-9nm)$|$Co(2.5nm)$|$AlO$_x$(2nm) and Ta(1-9nm)$| $Co(2.5nm)$|$AlO$_x$(2nm) layers were grown by dc magnetron sputtering on thermally oxidized Si wafers. The deposition rates were 0.185~nm/s for Pt, 0.067~nm/s for Ta, 0.052~nm/s for Co, and 0.077~nm/s for Al. The deposition pressure was 2~mTorr and the power was 50~W for all targets. The Al capping layers were oxidized in-situ by a 7~mTorr O$_2$ plasma at 10~W during 35~s. The layers were subsequently patterned into 6-terminals Hall bars by using standard optical lithography and Ar milling procedures. The Hall bar dimensions are $w$ for the current line width, $w/2$ for the Hall branch width, and $l=4w$ is the distance between two Hall branches, where $w$ varies between 4 and 10~$\mu$m.\\
\textbf{Characterization.} All layers possess spontaneous isotropic in-plane magnetization. To determine the saturation magnetization of Co we have performed anomalous Hall effect measurements on an 8~nm-thick Co reference sample with $B_{ext}\parallel \mathbf{z}$. The field required to fully saturate $\mathbf{M}$ out-of-plane is about 1.5~T, which, assuming that perpendicular magnetic anisotropy is negligible in this layer, is close to $\mu_0 M_s$ expected of Co. Similar measurements on Ta(6nm)$|$Co(2.5nm) and Pt(6nm)$|$Co(2.5nm) layers give saturation fields of 1.45~T and 0.8~T, respectively. This is attributed to the small (large) perpendicular interface anisotropy contribution of the Ta$|$Co (Pt$|$Co) interface, reducing the field required to saturate the magnetization out-of-plane.  4-point resistivity measurements on single Co(8nm), Ta(6nm) and Pt(6nm) layers yield $\rho_{Co}=25.3~\mu\Omega$cm, $\rho_{Ta}=237~\mu\Omega$cm and $\rho_{Pt}=34.1~\mu\Omega$cm, in line with expectations for Pt and Co thin films, and the $\beta$-phase of Ta. The magnetoresistance and Hall voltage measurements were performed at room temperature by using an ac current $I=I_0 \sin \omega t$ where $\omega / 2\pi=10$~Hz, generated by a Keithley-6221 constant current source. For the data reported in Fig.~2 the peak amplitude of the injected ac current was set to 8.5~mA, corresponding to a nominal current density of $j=10^7$~A/cm$^2$. In other measurements with different device size and thickness, the current was adapted to have the same current density. The longitudinal and transverse voltages were recorded simultaneously by using a 24-bit resolution National Instruments PXI-4462 dynamic signal analyser, during 10~s at each angle position in a uniform external field of 1.7~T. The rotation of the sample was provided by a motorized stage having a precision of 0.02 degrees. The acquired voltages were fast Fourier transformed (FFT) to extract the first and second harmonic voltage signals $V_{\omega}$ and $V_{2\omega}$. The corresponding resistances are given by $R_{\omega}=V_{\omega}/I_0$ and $R_{2\omega}=V_{2\omega}/I_0$ (peak values).

\bigskip \noindent \textbf{Acknowledgments}\\
This work was funded by the Swiss National Science Foundation (Grant No. 200021-153404) and the European Commission under the $7^{th}$ Framework Program (SPOT project, Grant No. 318144).

\bigskip \noindent \textbf{Author contributions}\\
C.O.A., K.G., and P.G. planned the experiments; M.G., A.G., S.F.A., and C.O.A. carried out the sample growth and patterning; C.O.A., K.G., and A.G. performed the measurements; C.O.A. and P.G. analyzed the data and wrote the manuscript. All authors contributed to the discussion of the data in the manuscript and Supplementary Information.

\bigskip \noindent \textbf{Additional information}\\
Correspondence and requests for materials should be addressed to \\
C.O.A. (\verb"can.onur.avci@mat.ethz.ch") and P.G. (\verb"pietro.gambardella@mat.ethz.ch").


\begin{thebibliography}{48}%
\makeatletter
\providecommand \@ifxundefined [1]{%
 \@ifx{#1\undefined}
}%
\providecommand \@ifnum [1]{%
 \ifnum #1\expandafter \@firstoftwo
 \else \expandafter \@secondoftwo
 \fi
}%
\providecommand \@ifx [1]{%
 \ifx #1\expandafter \@firstoftwo
 \else \expandafter \@secondoftwo
 \fi
}%
\providecommand \natexlab [1]{#1}%
\providecommand \enquote  [1]{``#1''}%
\providecommand \bibnamefont  [1]{#1}%
\providecommand \bibfnamefont [1]{#1}%
\providecommand \citenamefont [1]{#1}%
\providecommand \href@noop [0]{\@secondoftwo}%
\providecommand \href [0]{\begingroup \@sanitize@url \@href}%
\providecommand \@href[1]{\@@startlink{#1}\@@href}%
\providecommand \@@href[1]{\endgroup#1\@@endlink}%
\providecommand \@sanitize@url [0]{\catcode `\\12\catcode `\$12\catcode
  `\&12\catcode `\#12\catcode `\^12\catcode `\_12\catcode `\%12\relax}%
\providecommand \@@startlink[1]{}%
\providecommand \@@endlink[0]{}%
\providecommand \url  [0]{\begingroup\@sanitize@url \@url }%
\providecommand \@url [1]{\endgroup\@href {#1}{\urlprefix }}%
\providecommand \urlprefix  [0]{URL }%
\providecommand \Eprint [0]{\href }%
\providecommand \doibase [0]{http://dx.doi.org/}%
\providecommand \selectlanguage [0]{\@gobble}%
\providecommand \bibinfo  [0]{\@secondoftwo}%
\providecommand \bibfield  [0]{\@secondoftwo}%
\providecommand \translation [1]{[#1]}%
\providecommand \BibitemOpen [0]{}%
\providecommand \bibitemStop [0]{}%
\providecommand \bibitemNoStop [0]{.\EOS\space}%
\providecommand \EOS [0]{\spacefactor3000\relax}%
\providecommand \BibitemShut  [1]{\csname bibitem#1\endcsname}%
\let\auto@bib@innerbib\@empty
\bibitem [{\citenamefont {Thomson}(1856)}]{Thomson1856prsl}%
  \BibitemOpen
  \bibfield  {author} {\bibinfo {author} {\bibfnamefont {W.}~\bibnamefont
  {Thomson}},\ }\href {\doibase 10.1098/rspl.1856.0144} {\bibfield  {journal}
  {\bibinfo  {journal} {Proc. R. Soc. London}\ }\textbf {\bibinfo {volume}
  {8}},\ \bibinfo {pages} {546} (\bibinfo {year} {1856})}\BibitemShut {NoStop}%
\bibitem [{\citenamefont {Campbell}\ \emph {et~al.}(1970)\citenamefont
  {Campbell}, \citenamefont {Fert},\ and\ \citenamefont
  {Jaoul}}]{Campbell1970jpc}%
  \BibitemOpen
  \bibfield  {author} {\bibinfo {author} {\bibfnamefont {I.}~\bibnamefont
  {Campbell}}, \bibinfo {author} {\bibfnamefont {A.}~\bibnamefont {Fert}}, \
  and\ \bibinfo {author} {\bibfnamefont {O.}~\bibnamefont {Jaoul}},\
  }\href@noop {} {\bibfield  {journal} {\bibinfo  {journal} {J. Phys. C}\
  }\textbf {\bibinfo {volume} {3}},\ \bibinfo {pages} {S95} (\bibinfo {year}
  {1970})}\BibitemShut {NoStop}%
\bibitem [{\citenamefont {McGuire}\ and\ \citenamefont
  {Potter}(1975)}]{Mcguire1975ieeetm}%
  \BibitemOpen
  \bibfield  {author} {\bibinfo {author} {\bibfnamefont {T.}~\bibnamefont
  {McGuire}}\ and\ \bibinfo {author} {\bibfnamefont {R.}~\bibnamefont
  {Potter}},\ }\href@noop {} {\bibfield  {journal} {\bibinfo  {journal} {IEEE
  Trans. Magn.}\ }\textbf {\bibinfo {volume} {11}},\ \bibinfo {pages} {1018}
  (\bibinfo {year} {1975})}\BibitemShut {NoStop}%
\bibitem [{\citenamefont {Baibich}\ \emph {et~al.}(1988)\citenamefont
  {Baibich}, \citenamefont {Broto}, \citenamefont {Fert}, \citenamefont
  {Van~Dau}, \citenamefont {Petroff}, \citenamefont {Etienne}, \citenamefont
  {Creuzet}, \citenamefont {Friederich},\ and\ \citenamefont
  {Chazelas}}]{BaibichPRL1988Gmr}%
  \BibitemOpen
  \bibfield  {author} {\bibinfo {author} {\bibfnamefont {M.~N.}\ \bibnamefont
  {Baibich}}, \bibinfo {author} {\bibfnamefont {J.~M.}\ \bibnamefont {Broto}},
  \bibinfo {author} {\bibfnamefont {A.}~\bibnamefont {Fert}}, \bibinfo {author}
  {\bibfnamefont {F.~N.}\ \bibnamefont {Van~Dau}}, \bibinfo {author}
  {\bibfnamefont {F.}~\bibnamefont {Petroff}}, \bibinfo {author} {\bibfnamefont
  {P.}~\bibnamefont {Etienne}}, \bibinfo {author} {\bibfnamefont
  {G.}~\bibnamefont {Creuzet}}, \bibinfo {author} {\bibfnamefont
  {A.}~\bibnamefont {Friederich}}, \ and\ \bibinfo {author} {\bibfnamefont
  {J.}~\bibnamefont {Chazelas}},\ }\href {\doibase 10.1103/PhysRevLett.61.2472}
  {\bibfield  {journal} {\bibinfo  {journal} {Phys. Rev. Lett.}\ }\textbf
  {\bibinfo {volume} {61}},\ \bibinfo {pages} {2472} (\bibinfo {year}
  {1988})}\BibitemShut {NoStop}%
\bibitem [{\citenamefont {Binasch}\ \emph {et~al.}(1989)\citenamefont
  {Binasch}, \citenamefont {Gr{\"u}nberg}, \citenamefont {Saurenbach},\ and\
  \citenamefont {Zinn}}]{Grunberg89prb}%
  \BibitemOpen
  \bibfield  {author} {\bibinfo {author} {\bibfnamefont {G.}~\bibnamefont
  {Binasch}}, \bibinfo {author} {\bibfnamefont {P.}~\bibnamefont
  {Gr{\"u}nberg}}, \bibinfo {author} {\bibfnamefont {F.}~\bibnamefont
  {Saurenbach}}, \ and\ \bibinfo {author} {\bibfnamefont {W.}~\bibnamefont
  {Zinn}},\ }\href@noop {} {\bibfield  {journal} {\bibinfo  {journal} {Phys.
  Rev. B}\ }\textbf {\bibinfo {volume} {39}},\ \bibinfo {pages} {4828}
  (\bibinfo {year} {1989})}\BibitemShut {NoStop}%
\bibitem [{\citenamefont {Johnson}\ and\ \citenamefont
  {Silsbee}(1985)}]{JohnsonPRL1985}%
  \BibitemOpen
  \bibfield  {author} {\bibinfo {author} {\bibfnamefont {M.}~\bibnamefont
  {Johnson}}\ and\ \bibinfo {author} {\bibfnamefont {R.~H.}\ \bibnamefont
  {Silsbee}},\ }\href@noop {} {\bibfield  {journal} {\bibinfo  {journal} {Phys.
  Rev. Lett.}\ }\textbf {\bibinfo {volume} {55}},\ \bibinfo {pages} {1790}
  (\bibinfo {year} {1985})}\BibitemShut {NoStop}%
\bibitem [{\citenamefont {Jedema}\ \emph {et~al.}(2001)\citenamefont {Jedema},
  \citenamefont {Filip},\ and\ \citenamefont {Van~Wees}}]{Jedema2001N}%
  \BibitemOpen
  \bibfield  {author} {\bibinfo {author} {\bibfnamefont {F.}~\bibnamefont
  {Jedema}}, \bibinfo {author} {\bibfnamefont {A.}~\bibnamefont {Filip}}, \
  and\ \bibinfo {author} {\bibfnamefont {B.}~\bibnamefont {Van~Wees}},\
  }\href@noop {} {\bibfield  {journal} {\bibinfo  {journal} {Nature}\ }\textbf
  {\bibinfo {volume} {410}},\ \bibinfo {pages} {345} (\bibinfo {year}
  {2001})}\BibitemShut {NoStop}%
\bibitem [{\citenamefont {Johnson}\ and\ \citenamefont
  {Silsbee}(1987)}]{Johnson1987PRB}%
  \BibitemOpen
  \bibfield  {author} {\bibinfo {author} {\bibfnamefont {M.}~\bibnamefont
  {Johnson}}\ and\ \bibinfo {author} {\bibfnamefont {R.}~\bibnamefont
  {Silsbee}},\ }\href@noop {} {\bibfield  {journal} {\bibinfo  {journal} {Phys.
  Rev. B}\ }\textbf {\bibinfo {volume} {35}},\ \bibinfo {pages} {4959}
  (\bibinfo {year} {1987})}\BibitemShut {NoStop}%
\bibitem [{\citenamefont {Van~Son}\ \emph {et~al.}(1987)\citenamefont
  {Van~Son}, \citenamefont {Van~Kempen},\ and\ \citenamefont
  {Wyder}}]{vanSon1987PRL}%
  \BibitemOpen
  \bibfield  {author} {\bibinfo {author} {\bibfnamefont {P.}~\bibnamefont
  {Van~Son}}, \bibinfo {author} {\bibfnamefont {H.}~\bibnamefont {Van~Kempen}},
  \ and\ \bibinfo {author} {\bibfnamefont {P.}~\bibnamefont {Wyder}},\
  }\href@noop {} {\bibfield  {journal} {\bibinfo  {journal} {Phys. Rev. Lett.}\
  }\textbf {\bibinfo {volume} {58}},\ \bibinfo {pages} {2271} (\bibinfo {year}
  {1987})}\BibitemShut {NoStop}%
\bibitem [{\citenamefont {Valet}\ and\ \citenamefont
  {Fert}(1993)}]{Valet1993PRB}%
  \BibitemOpen
  \bibfield  {author} {\bibinfo {author} {\bibfnamefont {T.}~\bibnamefont
  {Valet}}\ and\ \bibinfo {author} {\bibfnamefont {A.}~\bibnamefont {Fert}},\
  }\href@noop {} {\bibfield  {journal} {\bibinfo  {journal} {Phys. Rev. B}\
  }\textbf {\bibinfo {volume} {48}},\ \bibinfo {pages} {7099} (\bibinfo {year}
  {1993})}\BibitemShut {NoStop}%
\bibitem [{\citenamefont {Maekawa}\ \emph {et~al.}(2012)\citenamefont
  {Maekawa}, \citenamefont {Valenzuela}, \citenamefont {Saitoh},\ and\
  \citenamefont {Kimura}}]{maekawa2012book}%
  \BibitemOpen
  \bibfield  {author} {\bibinfo {author} {\bibfnamefont {S.}~\bibnamefont
  {Maekawa}}, \bibinfo {author} {\bibfnamefont {S.~O.}\ \bibnamefont
  {Valenzuela}}, \bibinfo {author} {\bibfnamefont {E.}~\bibnamefont {Saitoh}},
  \ and\ \bibinfo {author} {\bibfnamefont {T.}~\bibnamefont {Kimura}},\
  }\href@noop {} {\emph {\bibinfo {title} {Spin Current}}}\ (\bibinfo
  {publisher} {Oxford University Press},\ \bibinfo {year} {2012})\BibitemShut
  {NoStop}%
\bibitem [{\citenamefont {Brataas}\ \emph {et~al.}(2006)\citenamefont
  {Brataas}, \citenamefont {Bauer},\ and\ \citenamefont
  {Kelly}}]{Brataas2006PR}%
  \BibitemOpen
  \bibfield  {author} {\bibinfo {author} {\bibfnamefont {A.}~\bibnamefont
  {Brataas}}, \bibinfo {author} {\bibfnamefont {G.~E.}\ \bibnamefont {Bauer}},
  \ and\ \bibinfo {author} {\bibfnamefont {P.~J.}\ \bibnamefont {Kelly}},\
  }\href@noop {} {\bibfield  {journal} {\bibinfo  {journal} {Phys. Rep.}\
  }\textbf {\bibinfo {volume} {427}},\ \bibinfo {pages} {157} (\bibinfo {year}
  {2006})}\BibitemShut {NoStop}%
\bibitem [{\citenamefont {Sinova}\ \emph {et~al.}()\citenamefont {Sinova},
  \citenamefont {Valenzuela}, \citenamefont {Wunderlich}, \citenamefont
  {Back},\ and\ \citenamefont {Jungwirth}}]{Sinova2014Arxiv}%
  \BibitemOpen
  \bibfield  {author} {\bibinfo {author} {\bibfnamefont {J.}~\bibnamefont
  {Sinova}}, \bibinfo {author} {\bibfnamefont {S.~O.}\ \bibnamefont
  {Valenzuela}}, \bibinfo {author} {\bibfnamefont {J.}~\bibnamefont
  {Wunderlich}}, \bibinfo {author} {\bibfnamefont {C.~H.}\ \bibnamefont
  {Back}}, \ and\ \bibinfo {author} {\bibfnamefont {T.}~\bibnamefont
  {Jungwirth}},\ }\href@noop {} {\bibinfo  {journal}
  {http://arxiv.org/abs/1411.3249}\ }\BibitemShut {NoStop}%
\bibitem [{\citenamefont {Miron}\ \emph {et~al.}(2011)\citenamefont {Miron},
  \citenamefont {Garello}, \citenamefont {Gaudin}, \citenamefont {Zermatten},
  \citenamefont {Costache}, \citenamefont {Auffret}, \citenamefont {Bandiera},
  \citenamefont {Rodmacq}, \citenamefont {Schuhl},\ and\ \citenamefont
  {Gambardella}}]{MironN2011}%
  \BibitemOpen
\bibfield  {journal} {  }\bibfield  {author} {\bibinfo {author} {\bibfnamefont
  {I.~M.}\ \bibnamefont {Miron}}, \bibinfo {author} {\bibfnamefont
  {K.}~\bibnamefont {Garello}}, \bibinfo {author} {\bibfnamefont
  {G.}~\bibnamefont {Gaudin}}, \bibinfo {author} {\bibfnamefont {P.-J.}\
  \bibnamefont {Zermatten}}, \bibinfo {author} {\bibfnamefont {M.~V.}\
  \bibnamefont {Costache}}, \bibinfo {author} {\bibfnamefont {S.}~\bibnamefont
  {Auffret}}, \bibinfo {author} {\bibfnamefont {S.}~\bibnamefont {Bandiera}},
  \bibinfo {author} {\bibfnamefont {B.}~\bibnamefont {Rodmacq}}, \bibinfo
  {author} {\bibfnamefont {A.}~\bibnamefont {Schuhl}}, \ and\ \bibinfo {author}
  {\bibfnamefont {P.}~\bibnamefont {Gambardella}},\ }\href@noop {} {\bibfield
  {journal} {\bibinfo  {journal} {Nature}\ }\textbf {\bibinfo {volume} {476}},\
  \bibinfo {pages} {189} (\bibinfo {year} {2011})}\BibitemShut {NoStop}%
\bibitem [{\citenamefont {Liu}\ \emph {et~al.}(2012)\citenamefont {Liu},
  \citenamefont {Pai}, \citenamefont {Li}, \citenamefont {Tseng}, \citenamefont
  {Ralph},\ and\ \citenamefont {Buhrman}}]{LiuS2012}%
  \BibitemOpen
  \bibfield  {author} {\bibinfo {author} {\bibfnamefont {L.}~\bibnamefont
  {Liu}}, \bibinfo {author} {\bibfnamefont {C.-F.}\ \bibnamefont {Pai}},
  \bibinfo {author} {\bibfnamefont {Y.}~\bibnamefont {Li}}, \bibinfo {author}
  {\bibfnamefont {H.}~\bibnamefont {Tseng}}, \bibinfo {author} {\bibfnamefont
  {D.}~\bibnamefont {Ralph}}, \ and\ \bibinfo {author} {\bibfnamefont
  {R.}~\bibnamefont {Buhrman}},\ }\href@noop {} {\bibfield  {journal} {\bibinfo
   {journal} {Science}\ }\textbf {\bibinfo {volume} {336}},\ \bibinfo {pages}
  {555} (\bibinfo {year} {2012})}\BibitemShut {NoStop}%
\bibitem [{\citenamefont {Garello}\ \emph {et~al.}(2013)\citenamefont
  {Garello}, \citenamefont {Miron}, \citenamefont {Avci}, \citenamefont
  {Freimuth}, \citenamefont {Mokrousov}, \citenamefont {Bl{\"u}gel},
  \citenamefont {Auffret}, \citenamefont {Boulle}, \citenamefont {Gaudin},\
  and\ \citenamefont {Gambardella}}]{GarelloNN2013}%
  \BibitemOpen
  \bibfield  {author} {\bibinfo {author} {\bibfnamefont {K.}~\bibnamefont
  {Garello}}, \bibinfo {author} {\bibfnamefont {I.~M.}\ \bibnamefont {Miron}},
  \bibinfo {author} {\bibfnamefont {C.~O.}\ \bibnamefont {Avci}}, \bibinfo
  {author} {\bibfnamefont {F.}~\bibnamefont {Freimuth}}, \bibinfo {author}
  {\bibfnamefont {Y.}~\bibnamefont {Mokrousov}}, \bibinfo {author}
  {\bibfnamefont {S.}~\bibnamefont {Bl{\"u}gel}}, \bibinfo {author}
  {\bibfnamefont {S.}~\bibnamefont {Auffret}}, \bibinfo {author} {\bibfnamefont
  {O.}~\bibnamefont {Boulle}}, \bibinfo {author} {\bibfnamefont
  {G.}~\bibnamefont {Gaudin}}, \ and\ \bibinfo {author} {\bibfnamefont
  {P.}~\bibnamefont {Gambardella}},\ }\href@noop {} {\bibfield  {journal}
  {\bibinfo  {journal} {Nature Nanotech.}\ }\textbf {\bibinfo {volume} {8}},\
  \bibinfo {pages} {587} (\bibinfo {year} {2013})}\BibitemShut {NoStop}%
\bibitem [{\citenamefont {Nakayama}\ \emph {et~al.}(2013)\citenamefont
  {Nakayama}, \citenamefont {Althammer}, \citenamefont {Chen}, \citenamefont
  {Uchida}, \citenamefont {Kajiwara}, \citenamefont {Kikuchi}, \citenamefont
  {Ohtani}, \citenamefont {Gepr{\"a}gs}, \citenamefont {Opel}, \citenamefont
  {Takahashi} \emph {et~al.}}]{NakayamaPRL2013}%
  \BibitemOpen
  \bibfield  {author} {\bibinfo {author} {\bibfnamefont {H.}~\bibnamefont
  {Nakayama}}, \bibinfo {author} {\bibfnamefont {M.}~\bibnamefont {Althammer}},
  \bibinfo {author} {\bibfnamefont {Y.-T.}\ \bibnamefont {Chen}}, \bibinfo
  {author} {\bibfnamefont {K.}~\bibnamefont {Uchida}}, \bibinfo {author}
  {\bibfnamefont {Y.}~\bibnamefont {Kajiwara}}, \bibinfo {author}
  {\bibfnamefont {D.}~\bibnamefont {Kikuchi}}, \bibinfo {author} {\bibfnamefont
  {T.}~\bibnamefont {Ohtani}}, \bibinfo {author} {\bibfnamefont
  {S.}~\bibnamefont {Gepr{\"a}gs}}, \bibinfo {author} {\bibfnamefont
  {M.}~\bibnamefont {Opel}}, \bibinfo {author} {\bibfnamefont {S.}~\bibnamefont
  {Takahashi}},  \emph {et~al.},\ }\href@noop {} {\bibfield  {journal}
  {\bibinfo  {journal} {Phys. Rev. Lett.}\ }\textbf {\bibinfo {volume} {110}},\
  \bibinfo {pages} {206601} (\bibinfo {year} {2013})}\BibitemShut {NoStop}%
\bibitem [{\citenamefont {Althammer}\ \emph {et~al.}(2013)\citenamefont
  {Althammer}, \citenamefont {Meyer}, \citenamefont {Nakayama}, \citenamefont
  {Schreier}, \citenamefont {Altmannshofer}, \citenamefont {Weiler},
  \citenamefont {Huebl}, \citenamefont {Gepr{\"a}gs}, \citenamefont {Opel},
  \citenamefont {Gross} \emph {et~al.}}]{althammer2013PRB}%
  \BibitemOpen
  \bibfield  {author} {\bibinfo {author} {\bibfnamefont {M.}~\bibnamefont
  {Althammer}}, \bibinfo {author} {\bibfnamefont {S.}~\bibnamefont {Meyer}},
  \bibinfo {author} {\bibfnamefont {H.}~\bibnamefont {Nakayama}}, \bibinfo
  {author} {\bibfnamefont {M.}~\bibnamefont {Schreier}}, \bibinfo {author}
  {\bibfnamefont {S.}~\bibnamefont {Altmannshofer}}, \bibinfo {author}
  {\bibfnamefont {M.}~\bibnamefont {Weiler}}, \bibinfo {author} {\bibfnamefont
  {H.}~\bibnamefont {Huebl}}, \bibinfo {author} {\bibfnamefont
  {S.}~\bibnamefont {Gepr{\"a}gs}}, \bibinfo {author} {\bibfnamefont
  {M.}~\bibnamefont {Opel}}, \bibinfo {author} {\bibfnamefont {R.}~\bibnamefont
  {Gross}},  \emph {et~al.},\ }\href@noop {} {\bibfield  {journal} {\bibinfo
  {journal} {Phys. Rev. B}\ }\textbf {\bibinfo {volume} {87}},\ \bibinfo
  {pages} {224401} (\bibinfo {year} {2013})}\BibitemShut {NoStop}%
\bibitem [{\citenamefont {Miao}\ \emph {et~al.}(2014)\citenamefont {Miao},
  \citenamefont {Huang}, \citenamefont {Qu},\ and\ \citenamefont
  {Chien}}]{Miao2014PRL}%
  \BibitemOpen
  \bibfield  {author} {\bibinfo {author} {\bibfnamefont {B.}~\bibnamefont
  {Miao}}, \bibinfo {author} {\bibfnamefont {S.}~\bibnamefont {Huang}},
  \bibinfo {author} {\bibfnamefont {D.}~\bibnamefont {Qu}}, \ and\ \bibinfo
  {author} {\bibfnamefont {C.}~\bibnamefont {Chien}},\ }\href@noop {}
  {\bibfield  {journal} {\bibinfo  {journal} {Phys. Rev. Lett.}\ }\textbf
  {\bibinfo {volume} {112}},\ \bibinfo {pages} {236601} (\bibinfo {year}
  {2014})}\BibitemShut {NoStop}%
\bibitem [{\citenamefont {Avci}\ \emph
  {et~al.}(2014{\natexlab{a}})\citenamefont {Avci}, \citenamefont {Garello},
  \citenamefont {Nistor}, \citenamefont {Godey}, \citenamefont {Ballesteros},
  \citenamefont {Mugarza}, \citenamefont {Barla}, \citenamefont {Valvidares},
  \citenamefont {Pellegrin}, \citenamefont {Ghosh}, \citenamefont {Miron},
  \citenamefont {Boulle}, \citenamefont {Auffret}, \citenamefont {Gaudin},\
  and\ \citenamefont {Gambardella}}]{AvciPRB2014}%
  \BibitemOpen
  \bibfield  {author} {\bibinfo {author} {\bibfnamefont {C.~O.}\ \bibnamefont
  {Avci}}, \bibinfo {author} {\bibfnamefont {K.}~\bibnamefont {Garello}},
  \bibinfo {author} {\bibfnamefont {C.}~\bibnamefont {Nistor}}, \bibinfo
  {author} {\bibfnamefont {S.}~\bibnamefont {Godey}}, \bibinfo {author}
  {\bibfnamefont {B.}~\bibnamefont {Ballesteros}}, \bibinfo {author}
  {\bibfnamefont {A.}~\bibnamefont {Mugarza}}, \bibinfo {author} {\bibfnamefont
  {A.}~\bibnamefont {Barla}}, \bibinfo {author} {\bibfnamefont
  {M.}~\bibnamefont {Valvidares}}, \bibinfo {author} {\bibfnamefont
  {E.}~\bibnamefont {Pellegrin}}, \bibinfo {author} {\bibfnamefont
  {A.}~\bibnamefont {Ghosh}}, \bibinfo {author} {\bibfnamefont {I.~M.}\
  \bibnamefont {Miron}}, \bibinfo {author} {\bibfnamefont {O.}~\bibnamefont
  {Boulle}}, \bibinfo {author} {\bibfnamefont {S.}~\bibnamefont {Auffret}},
  \bibinfo {author} {\bibfnamefont {G.}~\bibnamefont {Gaudin}}, \ and\ \bibinfo
  {author} {\bibfnamefont {P.}~\bibnamefont {Gambardella}},\ }\href@noop {}
  {\bibfield  {journal} {\bibinfo  {journal} {Phys. Rev. B}\ }\textbf {\bibinfo
  {volume} {89}},\ \bibinfo {pages} {214419} (\bibinfo {year}
  {2014}{\natexlab{a}})}\BibitemShut {NoStop}%
\bibitem [{\citenamefont {Hayashi}\ \emph {et~al.}(2014)\citenamefont
  {Hayashi}, \citenamefont {Kim}, \citenamefont {Yamanouchi},\ and\
  \citenamefont {Ohno}}]{HayashiPRB2014}%
  \BibitemOpen
  \bibfield  {author} {\bibinfo {author} {\bibfnamefont {M.}~\bibnamefont
  {Hayashi}}, \bibinfo {author} {\bibfnamefont {J.}~\bibnamefont {Kim}},
  \bibinfo {author} {\bibfnamefont {M.}~\bibnamefont {Yamanouchi}}, \ and\
  \bibinfo {author} {\bibfnamefont {H.}~\bibnamefont {Ohno}},\ }\href@noop {}
  {\bibfield  {journal} {\bibinfo  {journal} {Phys. Rev. B}\ }\textbf {\bibinfo
  {volume} {89}},\ \bibinfo {pages} {144425} (\bibinfo {year}
  {2014})}\BibitemShut {NoStop}%
\bibitem [{\citenamefont {Avci}\ \emph
  {et~al.}(2014{\natexlab{b}})\citenamefont {Avci}, \citenamefont {Garello},
  \citenamefont {Gabureac}, \citenamefont {Ghosh}, \citenamefont {Fuhrer},
  \citenamefont {Alvarado},\ and\ \citenamefont {Gambardella}}]{AvciPRB2014b}%
  \BibitemOpen
  \bibfield  {author} {\bibinfo {author} {\bibfnamefont {C.~O.}\ \bibnamefont
  {Avci}}, \bibinfo {author} {\bibfnamefont {K.}~\bibnamefont {Garello}},
  \bibinfo {author} {\bibfnamefont {M.}~\bibnamefont {Gabureac}}, \bibinfo
  {author} {\bibfnamefont {A.}~\bibnamefont {Ghosh}}, \bibinfo {author}
  {\bibfnamefont {A.}~\bibnamefont {Fuhrer}}, \bibinfo {author} {\bibfnamefont
  {S.~F.}\ \bibnamefont {Alvarado}}, \ and\ \bibinfo {author} {\bibfnamefont
  {P.}~\bibnamefont {Gambardella}},\ }\href {\doibase
  10.1103/PhysRevB.90.224427} {\bibfield  {journal} {\bibinfo  {journal} {Phys.
  Rev. B}\ }\textbf {\bibinfo {volume} {90}},\ \bibinfo {pages} {224427}
  (\bibinfo {year} {2014}{\natexlab{b}})}\BibitemShut {NoStop}%
\bibitem [{\citenamefont {Kobs}\ \emph {et~al.}(2011)\citenamefont {Kobs},
  \citenamefont {He{\ss}e}, \citenamefont {Kreuzpaintner}, \citenamefont
  {Winkler}, \citenamefont {Lott}, \citenamefont {Weinberger}, \citenamefont
  {Schreyer},\ and\ \citenamefont {Oepen}}]{KobsPRL2011}%
  \BibitemOpen
  \bibfield  {author} {\bibinfo {author} {\bibfnamefont {A.}~\bibnamefont
  {Kobs}}, \bibinfo {author} {\bibfnamefont {S.}~\bibnamefont {He{\ss}e}},
  \bibinfo {author} {\bibfnamefont {W.}~\bibnamefont {Kreuzpaintner}}, \bibinfo
  {author} {\bibfnamefont {G.}~\bibnamefont {Winkler}}, \bibinfo {author}
  {\bibfnamefont {D.}~\bibnamefont {Lott}}, \bibinfo {author} {\bibfnamefont
  {P.}~\bibnamefont {Weinberger}}, \bibinfo {author} {\bibfnamefont
  {A.}~\bibnamefont {Schreyer}}, \ and\ \bibinfo {author} {\bibfnamefont
  {H.}~\bibnamefont {Oepen}},\ }\href@noop {} {\bibfield  {journal} {\bibinfo
  {journal} {Phys. Rev. Lett.}\ }\textbf {\bibinfo {volume} {106}},\ \bibinfo
  {pages} {217207} (\bibinfo {year} {2011})}\BibitemShut {NoStop}%
\bibitem [{\citenamefont {Lu}\ \emph {et~al.}(2013)\citenamefont {Lu},
  \citenamefont {Cai}, \citenamefont {Huang}, \citenamefont {Qu}, \citenamefont
  {Miao},\ and\ \citenamefont {Chien}}]{LuPRB2013}%
  \BibitemOpen
  \bibfield  {author} {\bibinfo {author} {\bibfnamefont {Y.}~\bibnamefont
  {Lu}}, \bibinfo {author} {\bibfnamefont {J.}~\bibnamefont {Cai}}, \bibinfo
  {author} {\bibfnamefont {S.}~\bibnamefont {Huang}}, \bibinfo {author}
  {\bibfnamefont {D.}~\bibnamefont {Qu}}, \bibinfo {author} {\bibfnamefont
  {B.}~\bibnamefont {Miao}}, \ and\ \bibinfo {author} {\bibfnamefont
  {C.}~\bibnamefont {Chien}},\ }\href@noop {} {\bibfield  {journal} {\bibinfo
  {journal} {Phys. Rev. B}\ }\textbf {\bibinfo {volume} {87}},\ \bibinfo
  {pages} {220409} (\bibinfo {year} {2013})}\BibitemShut {NoStop}%
\bibitem [{\citenamefont {Hahn}\ \emph {et~al.}(2013)\citenamefont {Hahn},
  \citenamefont {De~Loubens}, \citenamefont {Klein}, \citenamefont {Viret},
  \citenamefont {Naletov},\ and\ \citenamefont {Youssef}}]{HahnPRB2013}%
  \BibitemOpen
  \bibfield  {author} {\bibinfo {author} {\bibfnamefont {C.}~\bibnamefont
  {Hahn}}, \bibinfo {author} {\bibfnamefont {G.}~\bibnamefont {De~Loubens}},
  \bibinfo {author} {\bibfnamefont {O.}~\bibnamefont {Klein}}, \bibinfo
  {author} {\bibfnamefont {M.}~\bibnamefont {Viret}}, \bibinfo {author}
  {\bibfnamefont {V.~V.}\ \bibnamefont {Naletov}}, \ and\ \bibinfo {author}
  {\bibfnamefont {J.~B.}\ \bibnamefont {Youssef}},\ }\href@noop {} {\bibfield
  {journal} {\bibinfo  {journal} {Phys. Rev. B}\ }\textbf {\bibinfo {volume}
  {87}},\ \bibinfo {pages} {174417} (\bibinfo {year} {2013})}\BibitemShut
  {NoStop}%
\bibitem [{\citenamefont {Kim}\ \emph {et~al.}(2013)\citenamefont {Kim},
  \citenamefont {Sinha}, \citenamefont {Hayashi}, \citenamefont {Yamanouchi},
  \citenamefont {Fukami}, \citenamefont {Suzuki}, \citenamefont {Mitani},\ and\
  \citenamefont {Ohno}}]{KimNM2013}%
  \BibitemOpen
  \bibfield  {author} {\bibinfo {author} {\bibfnamefont {J.}~\bibnamefont
  {Kim}}, \bibinfo {author} {\bibfnamefont {J.}~\bibnamefont {Sinha}}, \bibinfo
  {author} {\bibfnamefont {M.}~\bibnamefont {Hayashi}}, \bibinfo {author}
  {\bibfnamefont {M.}~\bibnamefont {Yamanouchi}}, \bibinfo {author}
  {\bibfnamefont {S.}~\bibnamefont {Fukami}}, \bibinfo {author} {\bibfnamefont
  {T.}~\bibnamefont {Suzuki}}, \bibinfo {author} {\bibfnamefont
  {S.}~\bibnamefont {Mitani}}, \ and\ \bibinfo {author} {\bibfnamefont
  {H.}~\bibnamefont {Ohno}},\ }\href@noop {} {\bibfield  {journal} {\bibinfo
  {journal} {Nature Mater.}\ }\textbf {\bibinfo {volume} {12}},\ \bibinfo
  {pages} {240} (\bibinfo {year} {2013})}\BibitemShut {NoStop}%
\bibitem [{\citenamefont {Weiler}\ \emph {et~al.}(2012)\citenamefont {Weiler},
  \citenamefont {Althammer}, \citenamefont {Czeschka}, \citenamefont {Huebl},
  \citenamefont {Wagner}, \citenamefont {Opel}, \citenamefont {Imort},
  \citenamefont {Reiss}, \citenamefont {Thomas}, \citenamefont {Gross} \emph
  {et~al.}}]{WeilerPRL2012}%
  \BibitemOpen
  \bibfield  {author} {\bibinfo {author} {\bibfnamefont {M.}~\bibnamefont
  {Weiler}}, \bibinfo {author} {\bibfnamefont {M.}~\bibnamefont {Althammer}},
  \bibinfo {author} {\bibfnamefont {F.~D.}\ \bibnamefont {Czeschka}}, \bibinfo
  {author} {\bibfnamefont {H.}~\bibnamefont {Huebl}}, \bibinfo {author}
  {\bibfnamefont {M.~S.}\ \bibnamefont {Wagner}}, \bibinfo {author}
  {\bibfnamefont {M.}~\bibnamefont {Opel}}, \bibinfo {author} {\bibfnamefont
  {I.-M.}\ \bibnamefont {Imort}}, \bibinfo {author} {\bibfnamefont
  {G.}~\bibnamefont {Reiss}}, \bibinfo {author} {\bibfnamefont
  {A.}~\bibnamefont {Thomas}}, \bibinfo {author} {\bibfnamefont
  {R.}~\bibnamefont {Gross}},  \emph {et~al.},\ }\href@noop {} {\bibfield
  {journal} {\bibinfo  {journal} {Phys. Rev. Lett.}\ }\textbf {\bibinfo
  {volume} {108}},\ \bibinfo {pages} {106602} (\bibinfo {year}
  {2012})}\BibitemShut {NoStop}%
\bibitem [{\citenamefont {Kikkawa}\ \emph {et~al.}(2013)\citenamefont
  {Kikkawa}, \citenamefont {Uchida}, \citenamefont {Shiomi}, \citenamefont
  {Qiu}, \citenamefont {Hou}, \citenamefont {Tian}, \citenamefont {Nakayama},
  \citenamefont {Jin},\ and\ \citenamefont {Saitoh}}]{KikkawaPRL2013}%
  \BibitemOpen
  \bibfield  {author} {\bibinfo {author} {\bibfnamefont {T.}~\bibnamefont
  {Kikkawa}}, \bibinfo {author} {\bibfnamefont {K.}~\bibnamefont {Uchida}},
  \bibinfo {author} {\bibfnamefont {Y.}~\bibnamefont {Shiomi}}, \bibinfo
  {author} {\bibfnamefont {Z.}~\bibnamefont {Qiu}}, \bibinfo {author}
  {\bibfnamefont {D.}~\bibnamefont {Hou}}, \bibinfo {author} {\bibfnamefont
  {D.}~\bibnamefont {Tian}}, \bibinfo {author} {\bibfnamefont {H.}~\bibnamefont
  {Nakayama}}, \bibinfo {author} {\bibfnamefont {X.-F.}\ \bibnamefont {Jin}}, \
  and\ \bibinfo {author} {\bibfnamefont {E.}~\bibnamefont {Saitoh}},\
  }\href@noop {} {\bibfield  {journal} {\bibinfo  {journal} {Phys. Rev. Lett.}\
  }\textbf {\bibinfo {volume} {110}},\ \bibinfo {pages} {067207} (\bibinfo
  {year} {2013})}\BibitemShut {NoStop}%
\bibitem [{\citenamefont {Camley}\ and\ \citenamefont
  {Barna{\'s}}(1989)}]{CamleyPRL1989}%
  \BibitemOpen
  \bibfield  {author} {\bibinfo {author} {\bibfnamefont {R.~E.}\ \bibnamefont
  {Camley}}\ and\ \bibinfo {author} {\bibfnamefont {J.}~\bibnamefont
  {Barna{\'s}}},\ }\href@noop {} {\bibfield  {journal} {\bibinfo  {journal}
  {Phys. Rev. Lett.}\ }\textbf {\bibinfo {volume} {63}},\ \bibinfo {pages}
  {664} (\bibinfo {year} {1989})}\BibitemShut {NoStop}%
\bibitem [{\citenamefont {Hood}\ and\ \citenamefont
  {Falicov}(1992)}]{HoodPRB1992}%
  \BibitemOpen
  \bibfield  {author} {\bibinfo {author} {\bibfnamefont {R.~Q.}\ \bibnamefont
  {Hood}}\ and\ \bibinfo {author} {\bibfnamefont {L.}~\bibnamefont {Falicov}},\
  }\href@noop {} {\bibfield  {journal} {\bibinfo  {journal} {Phys. Rev. B}\
  }\textbf {\bibinfo {volume} {46}},\ \bibinfo {pages} {8287} (\bibinfo {year}
  {1992})}\BibitemShut {NoStop}%
\bibitem [{\citenamefont {Zhang}\ \emph {et~al.}(2013)\citenamefont {Zhang},
  \citenamefont {Vlaminck}, \citenamefont {Pearson}, \citenamefont {Divan},
  \citenamefont {Bader},\ and\ \citenamefont {Hoffmann}}]{ZhangAPL2013}%
  \BibitemOpen
  \bibfield  {author} {\bibinfo {author} {\bibfnamefont {W.}~\bibnamefont
  {Zhang}}, \bibinfo {author} {\bibfnamefont {V.}~\bibnamefont {Vlaminck}},
  \bibinfo {author} {\bibfnamefont {J.~E.}\ \bibnamefont {Pearson}}, \bibinfo
  {author} {\bibfnamefont {R.}~\bibnamefont {Divan}}, \bibinfo {author}
  {\bibfnamefont {S.~D.}\ \bibnamefont {Bader}}, \ and\ \bibinfo {author}
  {\bibfnamefont {A.}~\bibnamefont {Hoffmann}},\ }\href@noop {} {\bibfield
  {journal} {\bibinfo  {journal} {Appl. Phys. Lett.}\ }\textbf {\bibinfo
  {volume} {103}},\ \bibinfo {pages} {242414} (\bibinfo {year}
  {2013})}\BibitemShut {NoStop}%
\bibitem [{\citenamefont {Dyakonov}(2007)}]{DyakonovPRL2007}%
  \BibitemOpen
  \bibfield  {author} {\bibinfo {author} {\bibfnamefont {M.}~\bibnamefont
  {Dyakonov}},\ }\href@noop {} {\bibfield  {journal} {\bibinfo  {journal}
  {Phys. Rev. Lett.}\ }\textbf {\bibinfo {volume} {99}},\ \bibinfo {pages}
  {126601} (\bibinfo {year} {2007})}\BibitemShut {NoStop}%
\bibitem [{\citenamefont {Dieny}(1992)}]{dieny92jpcm}%
  \BibitemOpen
  \bibfield  {author} {\bibinfo {author} {\bibfnamefont {B.}~\bibnamefont
  {Dieny}},\ }\href@noop {} {\bibfield  {journal} {\bibinfo  {journal} {J.
  Phys. Condens. Matter}\ }\textbf {\bibinfo {volume} {4}},\ \bibinfo {pages}
  {8009} (\bibinfo {year} {1992})}\BibitemShut {NoStop}%
\bibitem [{\citenamefont {Nguyen}\ \emph {et~al.}(2014)\citenamefont {Nguyen},
  \citenamefont {Pratt~Jr},\ and\ \citenamefont {Bass}}]{bass2014jmmm}%
  \BibitemOpen
  \bibfield  {author} {\bibinfo {author} {\bibfnamefont {H.}~\bibnamefont
  {Nguyen}}, \bibinfo {author} {\bibfnamefont {W.}~\bibnamefont {Pratt~Jr}}, \
  and\ \bibinfo {author} {\bibfnamefont {J.}~\bibnamefont {Bass}},\ }\href@noop
  {} {\bibfield  {journal} {\bibinfo  {journal} {J. Magn. Magn. Mat.}\ }\textbf
  {\bibinfo {volume} {361}},\ \bibinfo {pages} {30} (\bibinfo {year}
  {2014})}\BibitemShut {NoStop}%
\bibitem [{\citenamefont {Valenzuela}\ and\ \citenamefont
  {Tinkham}(2006)}]{Valenzuela2006N}%
  \BibitemOpen
  \bibfield  {author} {\bibinfo {author} {\bibfnamefont {S.~O.}\ \bibnamefont
  {Valenzuela}}\ and\ \bibinfo {author} {\bibfnamefont {M.}~\bibnamefont
  {Tinkham}},\ }\href@noop {} {\bibfield  {journal} {\bibinfo  {journal}
  {Nature}\ }\textbf {\bibinfo {volume} {442}},\ \bibinfo {pages} {176}
  (\bibinfo {year} {2006})}\BibitemShut {NoStop}%
\bibitem [{\citenamefont {Zhang}\ \emph {et~al.}(1992)\citenamefont {Zhang},
  \citenamefont {Levy},\ and\ \citenamefont {Fert}}]{zhang1992prb}%
  \BibitemOpen
  \bibfield  {author} {\bibinfo {author} {\bibfnamefont {S.}~\bibnamefont
  {Zhang}}, \bibinfo {author} {\bibfnamefont {P.}~\bibnamefont {Levy}}, \ and\
  \bibinfo {author} {\bibfnamefont {A.}~\bibnamefont {Fert}},\ }\href@noop {}
  {\bibfield  {journal} {\bibinfo  {journal} {Phys. Rev. B}\ }\textbf {\bibinfo
  {volume} {45}},\ \bibinfo {pages} {8689} (\bibinfo {year}
  {1992})}\BibitemShut {NoStop}%
\bibitem [{\citenamefont {Haney}\ \emph {et~al.}(2013)\citenamefont {Haney},
  \citenamefont {Lee}, \citenamefont {Lee}, \citenamefont {Manchon},\ and\
  \citenamefont {Stiles}}]{HaneyPRB2013}%
  \BibitemOpen
  \bibfield  {author} {\bibinfo {author} {\bibfnamefont {P.~M.}\ \bibnamefont
  {Haney}}, \bibinfo {author} {\bibfnamefont {H.-W.}\ \bibnamefont {Lee}},
  \bibinfo {author} {\bibfnamefont {K.-J.}\ \bibnamefont {Lee}}, \bibinfo
  {author} {\bibfnamefont {A.}~\bibnamefont {Manchon}}, \ and\ \bibinfo
  {author} {\bibfnamefont {M.}~\bibnamefont {Stiles}},\ }\href@noop {}
  {\bibfield  {journal} {\bibinfo  {journal} {Phys. Rev. B}\ }\textbf {\bibinfo
  {volume} {87}},\ \bibinfo {pages} {174411} (\bibinfo {year}
  {2013})}\BibitemShut {NoStop}%
\bibitem [{\citenamefont {Manchon}\ and\ \citenamefont
  {Zhang}(2008)}]{ManchonPRB2008}%
  \BibitemOpen
  \bibfield  {author} {\bibinfo {author} {\bibfnamefont {A.}~\bibnamefont
  {Manchon}}\ and\ \bibinfo {author} {\bibfnamefont {S.}~\bibnamefont
  {Zhang}},\ }\href@noop {} {\bibfield  {journal} {\bibinfo  {journal} {Phys.
  Rev. B}\ }\textbf {\bibinfo {volume} {78}},\ \bibinfo {pages} {212405}
  (\bibinfo {year} {2008})}\BibitemShut {NoStop}%
\bibitem [{\citenamefont {Miron}\ \emph {et~al.}(2010)\citenamefont {Miron},
  \citenamefont {Gaudin}, \citenamefont {Auffret}, \citenamefont {Rodmacq},
  \citenamefont {Schuhl}, \citenamefont {Pizzini}, \citenamefont {Vogel},\ and\
  \citenamefont {Gambardella}}]{MironNM2010}%
  \BibitemOpen
  \bibfield  {author} {\bibinfo {author} {\bibfnamefont {I.~M.}\ \bibnamefont
  {Miron}}, \bibinfo {author} {\bibfnamefont {G.}~\bibnamefont {Gaudin}},
  \bibinfo {author} {\bibfnamefont {S.}~\bibnamefont {Auffret}}, \bibinfo
  {author} {\bibfnamefont {B.}~\bibnamefont {Rodmacq}}, \bibinfo {author}
  {\bibfnamefont {A.}~\bibnamefont {Schuhl}}, \bibinfo {author} {\bibfnamefont
  {S.}~\bibnamefont {Pizzini}}, \bibinfo {author} {\bibfnamefont
  {J.}~\bibnamefont {Vogel}}, \ and\ \bibinfo {author} {\bibfnamefont
  {P.}~\bibnamefont {Gambardella}},\ }\href@noop {} {\bibfield  {journal}
  {\bibinfo  {journal} {Nature Mater.}\ }\textbf {\bibinfo {volume} {9}},\
  \bibinfo {pages} {230} (\bibinfo {year} {2010})}\BibitemShut {NoStop}%
\bibitem [{\citenamefont {Mahfouzi}\ \emph {et~al.}(2012)\citenamefont
  {Mahfouzi}, \citenamefont {Nagaosa},\ and\ \citenamefont
  {Nikoli{\'c}}}]{mahfouzi2012prl}%
  \BibitemOpen
  \bibfield  {author} {\bibinfo {author} {\bibfnamefont {F.}~\bibnamefont
  {Mahfouzi}}, \bibinfo {author} {\bibfnamefont {N.}~\bibnamefont {Nagaosa}}, \
  and\ \bibinfo {author} {\bibfnamefont {B.~K.}\ \bibnamefont {Nikoli{\'c}}},\
  }\href@noop {} {\bibfield  {journal} {\bibinfo  {journal} {Phys. Rev. Lett.}\
  }\textbf {\bibinfo {volume} {109}},\ \bibinfo {pages} {166602} (\bibinfo
  {year} {2012})}\BibitemShut {NoStop}%
\bibitem [{\citenamefont {S{\'a}nchez}\ and\ \citenamefont
  {B{\"u}ttiker}(2004)}]{buttiker04prl}%
  \BibitemOpen
  \bibfield  {author} {\bibinfo {author} {\bibfnamefont {D.}~\bibnamefont
  {S{\'a}nchez}}\ and\ \bibinfo {author} {\bibfnamefont {M.}~\bibnamefont
  {B{\"u}ttiker}},\ }\href@noop {} {\bibfield  {journal} {\bibinfo  {journal}
  {Phys. Rev. Lett.}\ }\textbf {\bibinfo {volume} {93}},\ \bibinfo {pages}
  {106802} (\bibinfo {year} {2004})}\BibitemShut {NoStop}%
\bibitem [{\citenamefont {Rikken}\ \emph {et~al.}(2001)\citenamefont {Rikken},
  \citenamefont {F\"olling},\ and\ \citenamefont {Wyder}}]{wyder01prl}%
  \BibitemOpen
  \bibfield  {author} {\bibinfo {author} {\bibfnamefont {G.}~\bibnamefont
  {Rikken}}, \bibinfo {author} {\bibfnamefont {J.}~\bibnamefont {F\"olling}}, \
  and\ \bibinfo {author} {\bibfnamefont {P.}~\bibnamefont {Wyder}},\
  }\href@noop {} {\bibfield  {journal} {\bibinfo  {journal} {Phys. Rev. Lett.}\
  }\textbf {\bibinfo {volume} {87}},\ \bibinfo {pages} {236602} (\bibinfo
  {year} {2001})}\BibitemShut {NoStop}%
\bibitem [{\citenamefont {Pop}\ \emph {et~al.}(2014)\citenamefont {Pop},
  \citenamefont {Auban-Senzier}, \citenamefont {Canadell}, \citenamefont
  {Rikken},\ and\ \citenamefont {Avarvari}}]{pop14nc}%
  \BibitemOpen
  \bibfield  {author} {\bibinfo {author} {\bibfnamefont {F.}~\bibnamefont
  {Pop}}, \bibinfo {author} {\bibfnamefont {P.}~\bibnamefont {Auban-Senzier}},
  \bibinfo {author} {\bibfnamefont {E.}~\bibnamefont {Canadell}}, \bibinfo
  {author} {\bibfnamefont {G.~L.}\ \bibnamefont {Rikken}}, \ and\ \bibinfo
  {author} {\bibfnamefont {N.}~\bibnamefont {Avarvari}},\ }\href@noop {}
  {\bibfield  {journal} {\bibinfo  {journal} {Nature Comm.}\ }\textbf {\bibinfo
  {volume} {5}} (\bibinfo {year} {2014})}\BibitemShut {NoStop}%
\bibitem [{\citenamefont {Vera-Marun}\ \emph {et~al.}(2012)\citenamefont
  {Vera-Marun}, \citenamefont {Ranjan},\ and\ \citenamefont {van
  Wees}}]{vera2012NP}%
  \BibitemOpen
  \bibfield  {author} {\bibinfo {author} {\bibfnamefont {I.~J.}\ \bibnamefont
  {Vera-Marun}}, \bibinfo {author} {\bibfnamefont {V.}~\bibnamefont {Ranjan}},
  \ and\ \bibinfo {author} {\bibfnamefont {B.~J.}\ \bibnamefont {van Wees}},\
  }\href@noop {} {\bibfield  {journal} {\bibinfo  {journal} {Nat. Phys.}\
  }\textbf {\bibinfo {volume} {8}},\ \bibinfo {pages} {313} (\bibinfo {year}
  {2012})}\BibitemShut {NoStop}%
\bibitem [{\citenamefont {Slachter}\ \emph {et~al.}(2010)\citenamefont
  {Slachter}, \citenamefont {Bakker}, \citenamefont {Adam},\ and\ \citenamefont
  {van Wees}}]{slachter2010NP}%
  \BibitemOpen
  \bibfield  {author} {\bibinfo {author} {\bibfnamefont {A.}~\bibnamefont
  {Slachter}}, \bibinfo {author} {\bibfnamefont {F.~L.}\ \bibnamefont
  {Bakker}}, \bibinfo {author} {\bibfnamefont {J.-P.}\ \bibnamefont {Adam}}, \
  and\ \bibinfo {author} {\bibfnamefont {B.~J.}\ \bibnamefont {van Wees}},\
  }\href@noop {} {\bibfield  {journal} {\bibinfo  {journal} {Nat. Phys.}\
  }\textbf {\bibinfo {volume} {6}},\ \bibinfo {pages} {879} (\bibinfo {year}
  {2010})}\BibitemShut {NoStop}%
\bibitem [{\citenamefont {Aziz}\ \emph {et~al.}(2009)\citenamefont {Aziz},
  \citenamefont {Wessely}, \citenamefont {Ali}, \citenamefont {Edwards},
  \citenamefont {Marrows}, \citenamefont {Hickey},\ and\ \citenamefont
  {Blamire}}]{AzizPRL2009}%
  \BibitemOpen
  \bibfield  {author} {\bibinfo {author} {\bibfnamefont {A.}~\bibnamefont
  {Aziz}}, \bibinfo {author} {\bibfnamefont {O.}~\bibnamefont {Wessely}},
  \bibinfo {author} {\bibfnamefont {M.}~\bibnamefont {Ali}}, \bibinfo {author}
  {\bibfnamefont {D.}~\bibnamefont {Edwards}}, \bibinfo {author} {\bibfnamefont
  {C.}~\bibnamefont {Marrows}}, \bibinfo {author} {\bibfnamefont
  {B.}~\bibnamefont {Hickey}}, \ and\ \bibinfo {author} {\bibfnamefont
  {M.}~\bibnamefont {Blamire}},\ }\href@noop {} {\bibfield  {journal} {\bibinfo
   {journal} {Phys. Rev. Lett.}\ }\textbf {\bibinfo {volume} {103}},\ \bibinfo
  {pages} {237203} (\bibinfo {year} {2009})}\BibitemShut {NoStop}%
\bibitem [{\citenamefont {Mellnik}\ \emph {et~al.}(2014)\citenamefont
  {Mellnik}, \citenamefont {Lee}, \citenamefont {Richardella}, \citenamefont
  {Grab}, \citenamefont {Mintun}, \citenamefont {Fischer}, \citenamefont
  {Vaezi}, \citenamefont {Manchon}, \citenamefont {Kim}, \citenamefont
  {Samarth} \emph {et~al.}}]{ralph14Nature}%
  \BibitemOpen
  \bibfield  {author} {\bibinfo {author} {\bibfnamefont {A.}~\bibnamefont
  {Mellnik}}, \bibinfo {author} {\bibfnamefont {J.}~\bibnamefont {Lee}},
  \bibinfo {author} {\bibfnamefont {A.}~\bibnamefont {Richardella}}, \bibinfo
  {author} {\bibfnamefont {J.}~\bibnamefont {Grab}}, \bibinfo {author}
  {\bibfnamefont {P.}~\bibnamefont {Mintun}}, \bibinfo {author} {\bibfnamefont
  {M.}~\bibnamefont {Fischer}}, \bibinfo {author} {\bibfnamefont
  {A.}~\bibnamefont {Vaezi}}, \bibinfo {author} {\bibfnamefont
  {A.}~\bibnamefont {Manchon}}, \bibinfo {author} {\bibfnamefont {E.-A.}\
  \bibnamefont {Kim}}, \bibinfo {author} {\bibfnamefont {N.}~\bibnamefont
  {Samarth}},  \emph {et~al.},\ }\href@noop {} {\bibfield  {journal} {\bibinfo
  {journal} {Nature}\ }\textbf {\bibinfo {volume} {511}},\ \bibinfo {pages}
  {449} (\bibinfo {year} {2014})}\BibitemShut {NoStop}%
\bibitem [{\citenamefont {Fan}\ \emph {et~al.}(2014)\citenamefont {Fan},
  \citenamefont {Upadhyaya}, \citenamefont {Kou}, \citenamefont {Lang},
  \citenamefont {Takei}, \citenamefont {Wang}, \citenamefont {Tang},
  \citenamefont {He}, \citenamefont {Chang}, \citenamefont {Montazeri} \emph
  {et~al.}}]{fan2014NM}%
  \BibitemOpen
  \bibfield  {author} {\bibinfo {author} {\bibfnamefont {Y.}~\bibnamefont
  {Fan}}, \bibinfo {author} {\bibfnamefont {P.}~\bibnamefont {Upadhyaya}},
  \bibinfo {author} {\bibfnamefont {X.}~\bibnamefont {Kou}}, \bibinfo {author}
  {\bibfnamefont {M.}~\bibnamefont {Lang}}, \bibinfo {author} {\bibfnamefont
  {S.}~\bibnamefont {Takei}}, \bibinfo {author} {\bibfnamefont
  {Z.}~\bibnamefont {Wang}}, \bibinfo {author} {\bibfnamefont {J.}~\bibnamefont
  {Tang}}, \bibinfo {author} {\bibfnamefont {L.}~\bibnamefont {He}}, \bibinfo
  {author} {\bibfnamefont {L.-T.}\ \bibnamefont {Chang}}, \bibinfo {author}
  {\bibfnamefont {M.}~\bibnamefont {Montazeri}},  \emph {et~al.},\ }\href@noop
  {} {\bibfield  {journal} {\bibinfo  {journal} {Nat.Mater.}\ }\textbf
  {\bibinfo {volume} {13}},\ \bibinfo {pages} {699} (\bibinfo {year}
  {2014})}\BibitemShut {NoStop}%
\end{thebibliography}
%

\clearpage

\clearpage

\clearpage

\clearpage

\clearpage

\end{document}


\title{
\bigskip
\bigskip
SUPPLEMENTARY INFORMATION\\
\bigskip
\bigskip
Unidirectional spin Hall magnetoresistance in ferromagnet/normal metal bilayers
}

\author{Can Onur Avci}
\author{Kevin Garello}
\author{Abhijit Ghosh}
\author{Mihai Gabureac}
\author{Santos F. Alvarado}
\author{Pietro Gambardella}
\affiliation{\smallskip Department of Materials, ETH Z\"{u}rich, H\"{o}nggerbergring 64, CH-8093 Z\"{u}rich, Switzerland}
\maketitle

\setcounter{equation}{0}
\setcounter{figure}{0}
\setcounter{table}{0}
\setcounter{page}{1}
\makeatletter
\renewcommand{\thesection}{SI \arabic{section}}
\renewcommand{\theequation}{S\arabic{equation}}
\renewcommand{\thefigure}{S\arabic{figure}}
$\renewcommand{\bibnumfmt}[1]{[S#1]}
$\renewcommand{\citenumfont}[1]{S#1}

\bigskip
\bigskip
\tableofcontents{}
\bigskip
\bigskip
\clearpage

\section{Harmonic analysis of the longitudinal and Hall resistances}
\label{Harmonic_analysis}

The longitudinal resistance ($R$) and transverse (Hall) resistance ($R^{H}$) are measured by applying an ac current  $I=I_{0}\sin(\omega t)$ of constant amplitude $I_{0}$ and frequency $\omega/2\pi=10$~Hz and recording the ac longitudinal ($V$) and transverse ($V^{H}$) voltage, respectively. Ohm's law for a current-dependent resistance reads $V(I)=R(I)\cdot I_{0}\sin(\omega t)$. Assuming that the current-induced resistance changes are small with respect to the linear resistance of the sample, $R(0)$, we can expand $R$ and $V$ as follows:
\begin{equation}
R(I)=R(0)+\frac{dR}{dI}dI = R(0)+\frac{1}{2}I_{0}\frac{dR}{dI}\sin(2\omega t) \, ,
\label{time dep R}
\end{equation}
and
\begin{equation}
V(I)=I_{0}R(0)\sin(\omega t)+\frac{1}{2}I^2_{0}\frac{dR}{dI}\sin(2\omega t) \, ,
\label{harmonic_signal}
\end{equation}
where the longitudinal voltage consists of first and second harmonic terms that scale with $I_0$ and $I^2_0$, respectively.
Accordingly, we define the first and second harmonic resistance as $R_{\omega} = R(0)$ and $R_{2\omega} = \frac{1}{2}I_{0}\frac{dR}{dI}$. Analogue expansions apply to the first and second harmonic Hall resistances, $R^H_{\omega}$ and $R^H_{2\omega}$. The first harmonic terms can be written as
\begin{equation}
R_{\omega}=R^{z}+(R^{x}-R^{z})\sin^{2}\theta\cos^{2}\varphi+(R^{y}-R^{z})\sin^{2}\theta\sin^{2}\varphi \, ,
\label{f_MR_SI}
\end{equation}
\begin{equation}
R^H_{\omega}=R_{AHE}\cos\theta+R_{PHE}\sin^2\theta\sin(2\varphi) \, ,
\label{f_Hall_SI}
\end{equation}
where $\theta$ and $\varphi$ are the polar and azimuthal magnetization angles, $R^i$ is the longitudinal resistance measured when the magnetization is saturated parallel to the direction $i=x,y,z$, and $R_{AHE}$ and $R_{PHE}$ are the anomalous Hall and planar Hall coefficients, respectively. Note that Eqs.~\ref{f_MR_SI} and \ref{f_Hall_SI} simply represent the conventional magnetoresistance (MR) and Hall resistance. The second harmonic terms, on the other hand, include all the contributions to $V$ and $V^{H}$ that vary quadratically with the current, specifically: the MR and Hall resistance changes due to current-induced spin-orbit torques (SOT) and Oersted field~\cite{GarelloNN2013,AvciPRB2014,HayashiPRB2014,KimNM2013}, thermoelectric effects~\cite{Smith1967,AvciPRB2014b}, and nonlinear resistive terms.
We have previously studied the influence of the SOT and thermal gradients ($\nabla T$) on the transverse voltage, showing that the corresponding second harmonic Hall resistances, $R^{H, SOT}_{2\omega}$ and $R^{H, \nabla T}_{2\omega}$, can be separately measured due to their different symmetry and field dependence~\cite{AvciPRB2014b}. In this paper we show there is an additional nonlinear contribution that adds to the longitudinal resistance, which we call unidirectional spin Hall magnetoresistance (USMR). We thus have

\begin{equation}
\label{2f_MR_SI}
R_{2\omega}\propto I_0(R^{SOT}_{2\omega}+R^{\nabla T}_{2\omega}+R^{USMR}_{2\omega}) \, ,
\end{equation}
\begin{equation}
\label{2f_Hall_SI}
R^H_{2\omega}\propto I_0(R^{H, SOT}_{2\omega}+R^{H, \nabla T}_{2\omega}) \, .
\end{equation}

The SOT terms are due to the current-induced oscillations of the magnetization that modulate the MR and Hall resistance through the dependence of the angles $\theta$ and $\varphi$ on the current. These terms read, for the longitudinal and transverse case, as

\begin{equation}
\label{2f_SOT_MR_SI}
R^{SOT}_{2\omega}=I_0[(R^x-R^z)\cos^{2}\varphi+(R^y-R^z)\sin^{2}\varphi]\frac{d\sin^{2}\theta}{dI}+
I_0[(R^y-R^z)\sin^{2}\theta-(R^x-R^z)\sin^{2}\theta]\frac{d\sin^{2}\varphi}{dI} \, ,
\end{equation}
\begin{equation}
R^{H, SOT}_{2\omega}=I_0(R_{AHE}-2R_{PHE}\cos\theta\sin 2\varphi)\frac{d\cos\theta}{dI} + I_0 R_{PHE}\sin^2\theta\frac{d\sin 2\varphi}{dI} \, .
\end{equation}

The thermoelectric terms are due to Joule heating and corresponding quadratic increase of the sample temperature with current, which gives rise to temperature gradients~\cite{AvciPRB2014b}:
\begin{equation}
\label{2f_thermo}
\nabla T \propto I_0^{2}\sin^2(\omega t)R_{0}=\frac{1}{2}I_0^{2}[1-\cos(2\omega t)]R_{0} \, .
\end{equation}
Such thermal gradients can give rise to the anomalous Nernst\cite{SchmidPRL2013,WeilerPRL2012}, spin Seebeck\cite{UchidaAPL2010} and magneto-thermopower effects\cite{SchmidPRL2013} with distinct angular dependencies. Their functional form in the longitudinal and transverse geometries is given in Sect.~\ref{T_effects} for an arbitrary $\nabla T$.

Finally, the modulation of the ferromagnet/normal metal (FM/NM) interface resistance by the spin Hall effect (SHE) gives rise to a second harmonic term
\begin{equation}
\label{2f_USMR}
R^{USMR}_{2\omega}\propto I_0 \theta_{SH} M_y  \, ,
\end{equation}
where $\theta_{SH}$ is the spin Hall angle of the NM and $M_y$ the $y$ component of the magnetization of the FM (see Figure~1 of the main text for the definition of the spatial coordinates).

\section{Determination of the polar magnetization angle using the Hall resistance}
\label{M_angle}

\begin{figure}[h]
  \centering
  \includegraphics[width=12 cm]{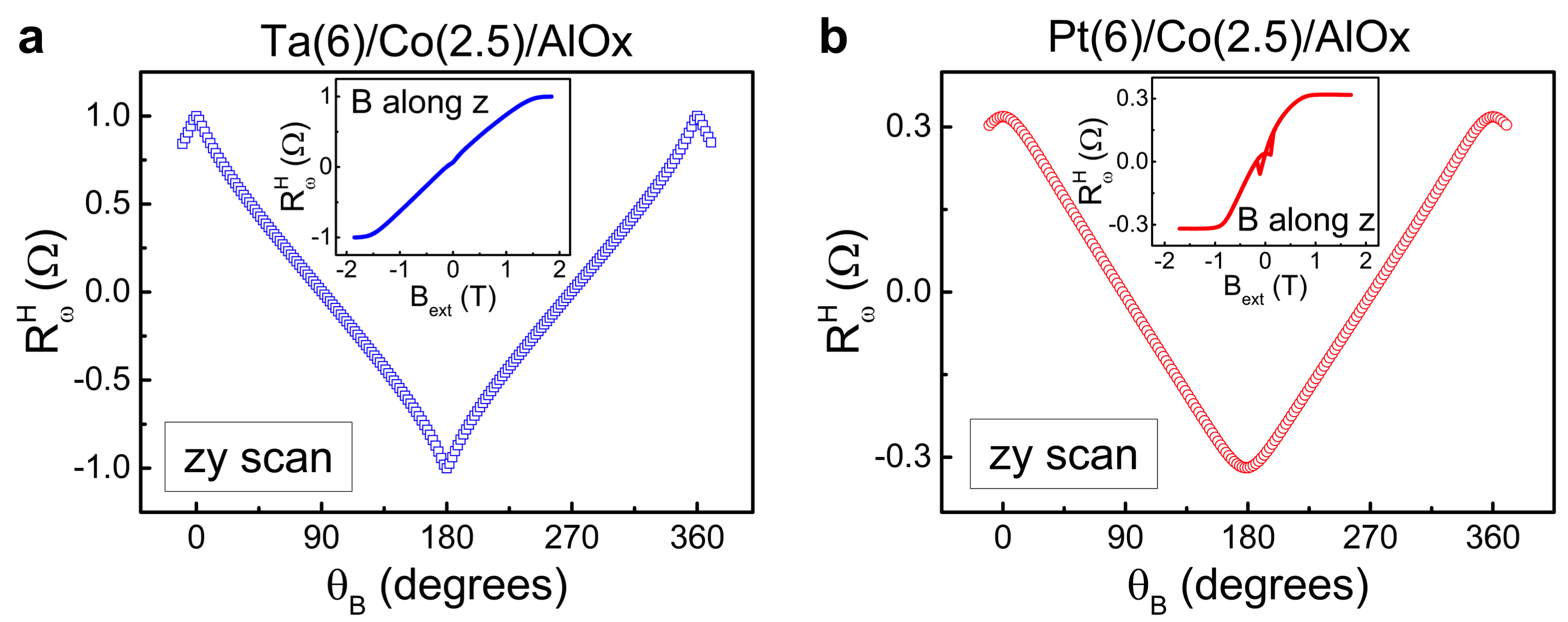}\\
  \caption{\textbf{Anomalous Hall resistance measurements.} \textbf{a,} $R^H_{\omega}=R_{AHE}\cos\theta$ of Ta(6nm)$|$Co(2.5nm) and \textbf{b,} Pt(6nm)$|$Co(2.5nm) measured during a $zy$ scan at fixed external field $B=1.7$~T. A small constant offset due to misalignment of the Hall branches has been subtracted from both curves. Insets: $R^H_{\omega}$ as a function of out-of-plane field. The features around $B_{ext}=0$~T are due to planar Hall effect contributions.}\label{fig1}
\end{figure}

If the magnetization is fully aligned to the external field, the polar and azimuthal angles of the $\mathbf{M}$ and $\mathbf{B}$ vectors coincide ($\theta=\theta_B$ and $\varphi=\varphi_B$). Even for magnetic fields of the order of a few Tesla, however, the demagnetizing field induces a significant deviation of $\theta$ from $\theta_B$, while, in the case of easy plane anisotropy of interest here, $\varphi=\varphi_B$ remains true. In order to fit the angular dependence of the MR in the $zx$, $zy$, and $xy$ planes using Eq.~\ref{f_MR_SI}, we therefore need an independent measurement of $\theta$. This is easily performed by measuring the Hall resistance as a function of $\theta_B$ in the $zx$ ($\varphi=0$) or $zy$ ($\varphi=\pi/2$) planes, as $R^H_{\omega}$ simplifies to $R^H_{\omega}=R_{AHE}\cos\theta$ in this case. The saturation value of the anomalous Hall resistance, $R_{AHE}$, is obtained as shown in Fig.~\ref{fig1}, where we report the Hall resistance for the $zy$ angular scans corresponding to (a) Ta(6nm)$|$Co(2.5nm) and (b) Pt(6nm)$|$Co(2.5nm), simultaneously recorded during the magnetoresistance (MR) measurements shown in Fig.~2 of the main text. A nearly identical $R^H_{\omega}$ is detected in the $zx$ plane for both samples. The insets show $R^H_{\omega}$ measured as a function of external field parallel to $z$, showing that the magnetization is technically saturated at a field of $\pm1.7$~T. Note that the saturation field of Ta$|$Co ($\sim 1.45$~T) is larger than that of Pt$|$Co ($\sim 0.8$~T) because of the strong perpendicular magnetic anisotropy (PMA) of Pt$|$Co, which counters the demagnetizing field of the Co layer.

\section{Influence of spin-orbit torques on the second harmonic MR measurements}
\label{SOT}

An electric current flowing in the plane of a FM/NM bilayer generates two qualitatively different types of SOT: a field-like (FL) torque $\boldsymbol{\tau}_{FL}\sim\mathbf{M}\times \hat{\mathbf{y}}$, and an antidamping-like (AD) torque $\boldsymbol{\tau}_{AD}\sim \mathbf{M}\times (\hat{\mathbf{y}} \times \mathbf{M})$, where $\hat{\mathbf{y}}$ is the in-plane axis perpendicular to the current flow direction $\hat{\mathbf{x}}$ (Refs.~\onlinecite{AndoPRL2008,MironNM2010,GarelloNN2013,KimNM2013,HaneyPRB2013}). The action of these torques is equivalent to that of two effective fields $\mathbf{B}_{FL}\sim \mathbf{M}\times \boldsymbol{\tau}_{FL}$ and $\mathbf{B}_{AD}\sim \mathbf{M}\times \boldsymbol{\tau}_{AD}$ perpendicular to the instantaneous direction of the magnetization .

Most of the studies using harmonic measurements so far have employed the Hall effect to reveal and quantify the SOT~\cite{PiAPL2010,GarelloNN2013,KimNM2013,AvciPRB2014,HayashiPRB2014}. However, SOT also give rise to second harmonic MR signals, $R^{SOT}_{2\omega}$, as noted in Ref.~\onlinecite{HayashiPRB2014}. Here we analyze the angular dependence of $R^{SOT}_{2\omega}$ based on symmetry arguments and macrospin simulations of the MR. To perform the simulations we compute the equilibrium position of $\mathbf{M}$ by considering the sum of all the torques acting on it: the torque due to the external field ($\boldsymbol{\tau}_B$), PMA ($\boldsymbol{\tau}_{PMA}$), demagnetizing field ($\boldsymbol{\tau}_{DEM}$), as well as the field-like SOT, including the torque due to the Oersted field, ($\boldsymbol{\tau}_{FL}$), and the antidamping-like SOT ($\boldsymbol{\tau}_{AD}$). At equilibrium, the position of the magnetization is defined by:

\begin{equation}
\label{allTorques}
\boldsymbol{\tau}_B+\boldsymbol{\tau}_{PMA}+\boldsymbol{\tau}_{DEM}+\boldsymbol{\tau}_{FL}+\boldsymbol{\tau}_{AD}=0
\end{equation}

\begin{figure}[t]
  \centering
  \includegraphics[width=17 cm]{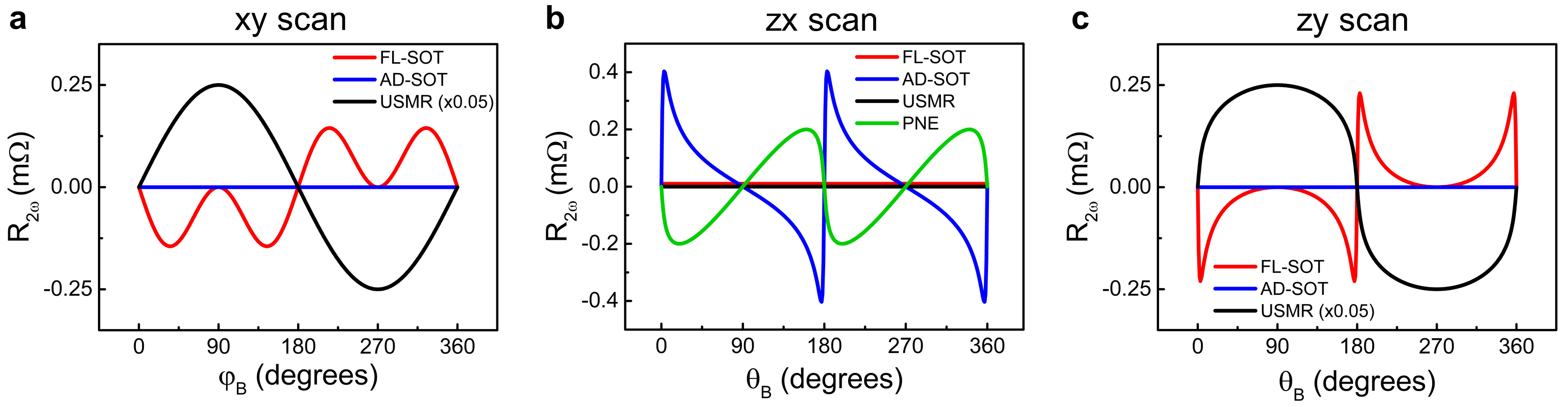}\\
  \caption{\textbf{Macrospin simulations of $R^{SOT}_{2\omega}$ and $R^{USMR}_{2\omega}$.} \textbf{a,} $xy$ scan, \textbf{b,} $zx$ scan, and \textbf{c,} $zy$ scan simulations. The simulated planar Nernst signal is also shown for comparison in \textbf{b}. Note that the SOT signals are approximately a factor 20 smaller compared to the USMR in the $xy$ and $zy$ plane. This is due to the fact that the magnetization becomes less susceptible to the current-induced torques when $\boldsymbol{\tau}_B$ is much larger compared to $\boldsymbol{\tau}_{AD}$ and $\boldsymbol{\tau}_{FL}$.}\label{fig2}

\end{figure}

We numerically solve the above equation for the equilibrium angles of the magnetization $\theta$ and $\varphi$ as a function of external field and current, using measured values of the following parameters for Ta(6nm)$|$Co(2.5nm): $R^x-R^y=1.36$~$\Omega$, $R^z-R^y=0.82$~$\Omega$, $R^H=1$~$\Omega$, $R^{PNE}=0.2$~m$\Omega$, $\boldsymbol{\tau}_{FL}=0.24$~mT, $\boldsymbol{\tau}_{AD}=0.6$~mT, $\boldsymbol{\tau}_{DEM}=1.45$~T, and $\boldsymbol{\tau}_{PMA} \approx 0$.
Figures~\ref{fig2}a-c compare the simulations of the second harmonic MR due to the SOT (Eq.~\ref{2f_SOT_MR_SI}) and USMR (Eq.~\ref{2f_USMR}) in the $xy$, $zx$, and $zy$ planes for an external field $B_{ext}=1.7$~T and current density $j=10^7$~A/cm$^2$.
The simulations show that the FL torque can give rise to a second harmonic MR contribution in the $xy$ and $zy$ scans whereas the contribution due to the AD torque is equally zero in both cases. In the $zx$ scan we have the opposite behavior, i.e., that the FL torque gives no contribution, whereas the AD torque gives rise to a non-zero signal. However, in all cases the MR contributions with SOT origin have a distinct symmetry with respect to the measurements reported in Fig.~2 of the main text. Moreover, $R^{SOT}_{2\omega}=0$ when either $\varphi_{B}$ or $\theta_{B}=\pi/2$ excluding the influence of SOT on the USMR measurements.

\section{Thermoelectric effects}
\label{T_effects}

Thermoelectric effects must be carefully considered in the measurement of electrical signals due to unavoidable Joule heating and consequent temperature gradients in the sample. The thermal gradient can have an arbitrary direction depending on the device geometry, stacking order of the layers, inhomogeneous current flow due to resistivity inhomogeneities in each device and give rise to a second harmonic signal, $R^{\nabla T}_{2\omega}$, due to their quadratic dependence on the injected current. A proper analysis incorporates the Nernst and Seebeck effects considering temperature gradients in all three directions in space. We have treated these effects case-by-case to identify their symmetry and their possible contribution to $R^{\nabla T}_{2\omega}$.

The anomalous Nernst effect (ANE) is analogous to the anomalous Hall effect driven by a temperature gradient~\cite{Smith1967,SchmidPRL2013}. The symmetry of the electric field driven by the ANE that gives rise to a second harmonic signal is as following:

\begin{equation}
\label{ANEsymm}
\mathbf{E}_{ANE}=-\alpha\mathbf{\nabla T} \times \mathbf{M} \, ,
\end{equation}
where $\alpha$ is the material dependent ANE coefficient. By considering an arbitrary temperature gradient $\mathbf{\nabla T}=(\nabla T_x,\nabla T_y,\nabla T_z)$ and magnetization direction $\hat{\mathbf{M}}=(\sin\theta\cos\varphi,\sin\theta\sin\varphi,\sin\theta)$ we find the following allowed symmetries for the second harmonic longitudinal ($R^{\nabla T}_{2\omega}$) and transverse ($R^{H, \nabla T}_{2\omega}$) resistances:

\begin{equation}
\label{R2fMR_ANE}
R^{ANE}_{2\omega}\propto \nabla T_y\cos\theta-\nabla T_z\sin\theta\sin\varphi \, ,
\end{equation}

\begin{equation}
\label{R2fHall_ANE}
R^{H, ANE}_{2\omega}\propto \nabla T_z\sin\theta\cos\varphi-\nabla T_x\cos\theta \, .
\end{equation}

The anisotropic magneto-thermopower (AMTEP) is the magnetization-dependent Seebeck effect and is the thermal analogous of the magnetoresistance~\cite{PuPRL2006}. Therefore, its manifestation follows Eq.~\ref{f_MR_SI} and, in terms of temperature gradients, reads:

\begin{equation}
\label{R2fMR_AMTEP}
R^{AMTEP}_{2\omega}\propto \nabla T_x(\sin^2\theta\cos^2\varphi+\sin^2\theta\sin^2\varphi) \, ,
\end{equation}

\begin{equation}
\label{R2fHall_AMTEP}
R^{H, AMTEP}_{2\omega}\propto \nabla T_y(\sin^2\theta\cos^2\varphi+\sin^2\theta\sin^2\varphi) \, .
\end{equation}

The transverse manifestation of the AMTEP, the so-called planar Nernst effect (PNE)~\cite{KyPSS1967,AveryPRL2012}, is analogous to the planar Hall effect and reads:

\begin{equation}
\label{R2fMR_PNE}
R^{PNE}_{2\omega}\propto \nabla T_y\sin^2\theta\sin2\varphi+\nabla T_z\sin2\theta\cos^2\varphi \, ,
\end{equation}

\begin{equation}
\label{R2fHall_PNE}
R^{H, PNE}_{2\omega}\propto \nabla T_x\sin^2\theta\sin2\varphi+\nabla T_z\sin2\theta\sin^2\varphi \, .
\end{equation}

In addition to these signals, in FM/NM bilayers one has to take into account the spin Seebeck effect (SSE), whereby a thermal gradient can drive a spin current that can be detected as a voltage across the bilayer via the inverse spin Hall effect. In our measurement geometry we could, in principle, observe the longitudinal SSE~\cite{UchidaAPL2010} due to a perpendicular thermal gradient, which would give rise to a signal with identical symmetry as that of the ANE (Eq.~\ref{R2fMR_ANE} and Eq.~\ref{R2fHall_ANE})~\cite{WeilerPRL2012}, although its sign would depend on the sign of the spin Hall effect in the NM layer.

\begin{figure}[h]
  \centering
  \includegraphics[width=12 cm]{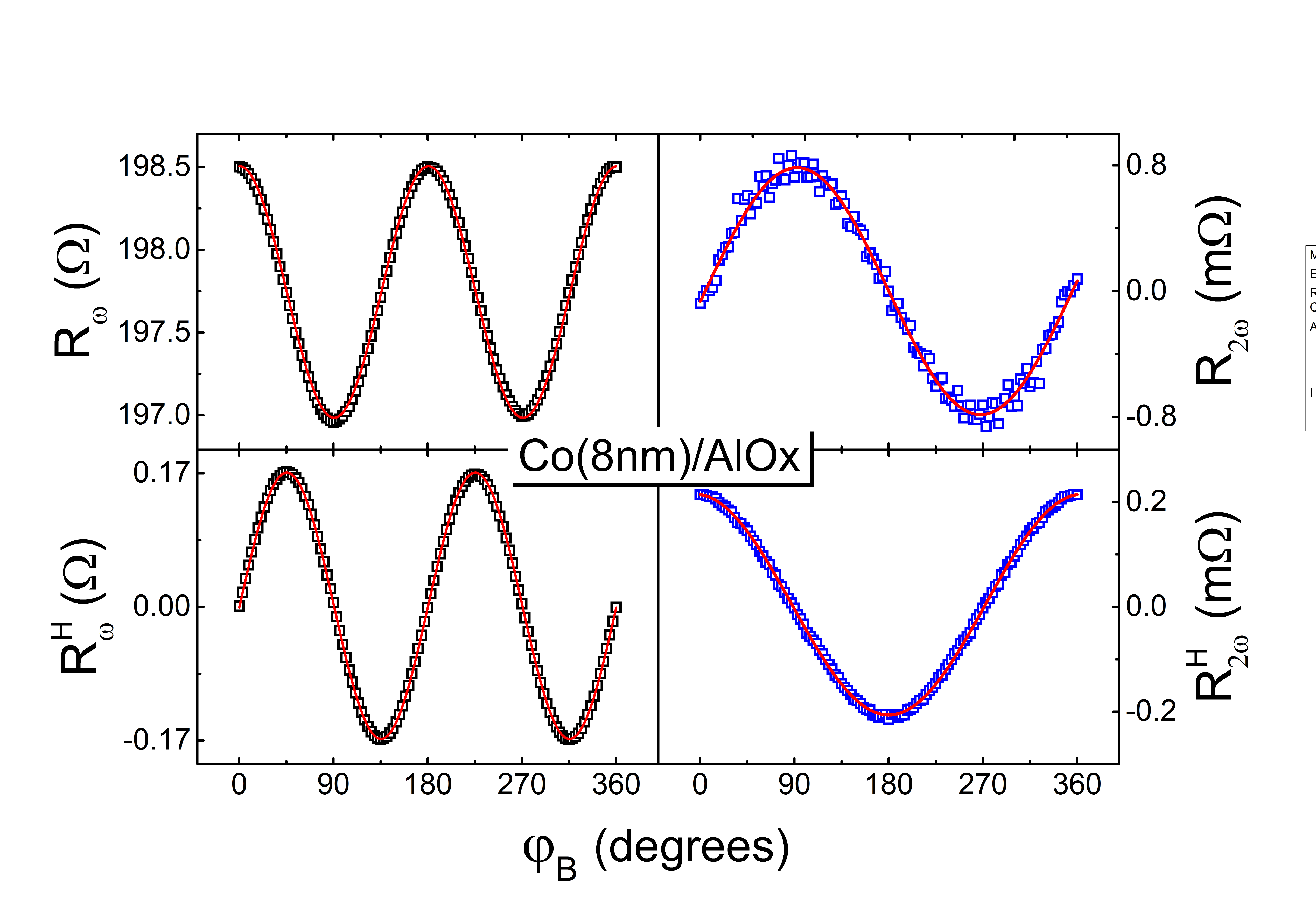}\\
  \caption{\textbf{Longitudinal and transverse measurements of the MR and ANE in a single Co layer.} First and second harmonic signals corresponding to the longitudinal and transverse resistance of Co(8nm). The sinusoidal fits (red curves) to the first harmonic signals show that the magnetization follows the applied field direction and the angular dependence of the MR ($R_{\omega}$) and planar Hall effect ($R^{H}_{\omega}$), whereas the second harmonic signals have the angular dependence expected of the ANE according to Eqs.~\ref{R2fMR_ANE} and \ref{R2fHall_ANE}. Note that $R_{2\omega}/R^{H}_{2\omega} = 4 = l/w$ as measured by scanning electron microscopy for this device.} \label{fig3}
\end{figure}

By considering Eq.~\ref{R2fMR_ANE} through Eq.~\ref{R2fHall_PNE} we notice that only a signal originating from the ANE (possibly including also the longitudinal SSE) and due to $\nabla T_z$ possesses the same symmetry as that of the USMR ($R^{USMR}_{2\omega} \propto \sin\theta\sin\varphi$). In the following, we discuss how these two effects can be separated. In previous work, we have shown that the ANE driven by a perpendicular temperature gradient induces a nonzero second harmonic Hall resistance in Co layers~\cite{AvciPRB2014b}. We use an 8~nm thick Co sample as a reference to reveal the influence of the ANE on the second harmonic longitudinal resistance. Figure~\ref{fig3} shows the first and second harmonic signals of longitudinal and transverse measurements when $\mathbf{M}$ is rotated in the $xy$ plane in a fixed field $B=0.4$~T. Sinusoidal fits (red solid curves) confirm the expected MR (planar Hall effect) symmetry in the longitudinal (transverse) first harmonic signals and the ANE symmetry in the second harmonic signals. Since the MR and planar Hall effect have the same microscopic origin, their amplitude is proportional to the physical distance over which they are measured. The same holds for the longitudinal and transverse ANE signals, which implies that the ratio between the longitudinal and transverse MR and ANE measurements is equal to the length/width ($l/w$) ratio of the Hall bar. This was verified experimentally by measuring the ratio $l/w$ of our Hall bars using scanning electron microscopy for devices of different size and including other reference layers such as Ti$|$Co bilayers for which the ANE and MR coefficients differ from those of a single Co layer~\cite{AvciPRB2014b}. Therefore, by measuring the thermal signal in the transverse geometry one can accurately determine its sign and magnitude in the longitudinal measurements.

\begin{figure}[h]
  \centering
  \includegraphics[width=8 cm]{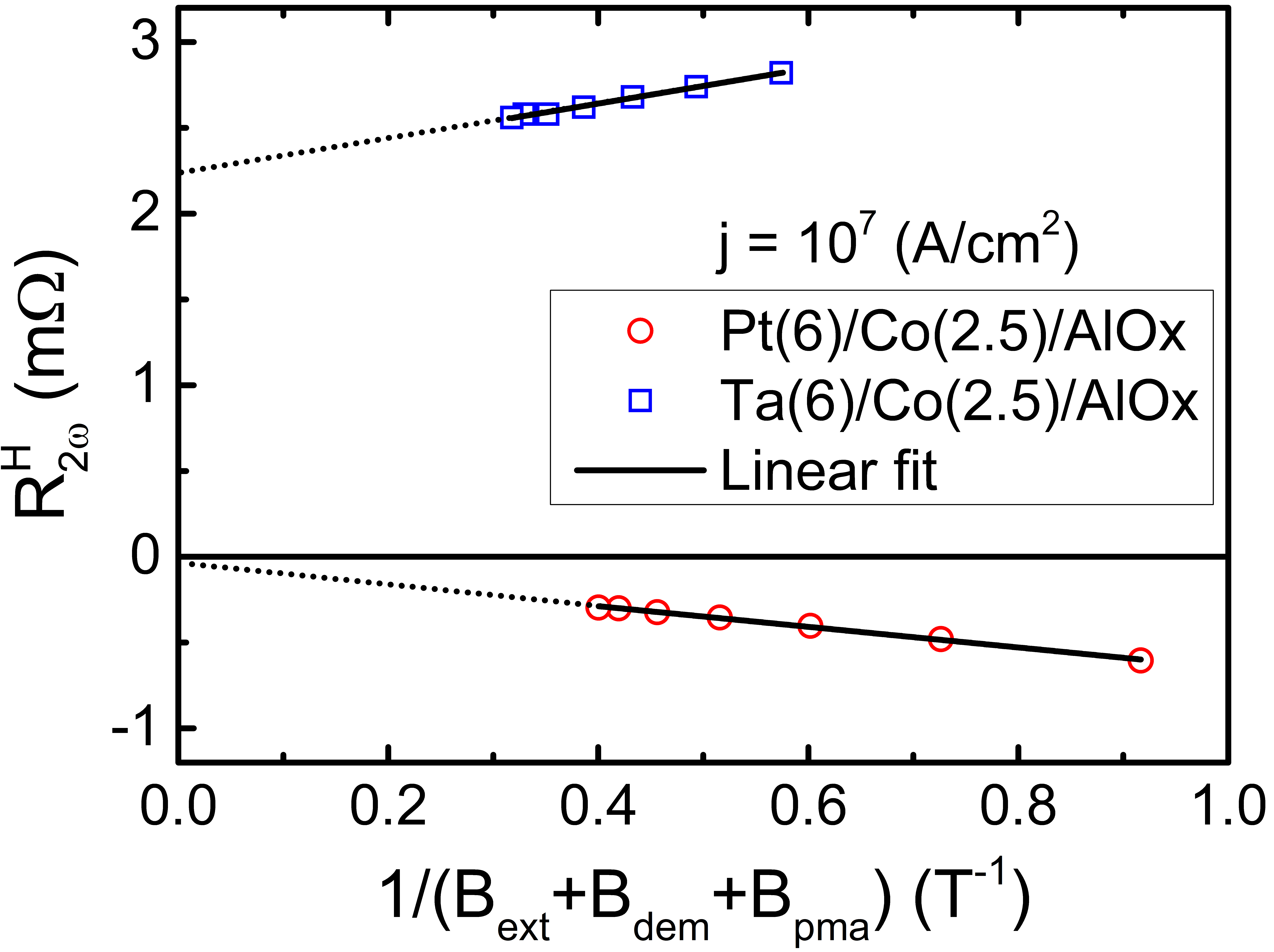}\\
  \caption{\textbf{Separation of AD-SOT and ANE contributions to $R^{H}_{2\omega}$.} Plot of the cosine component of $R^{H}_{2\omega}$ due to the AD-SOT and ANE as a function of $(B_{ext}+B_{DEM}+B_{PMA})^{-1}$ measured in an $xy$ scan for Ta(6nm)$|$Co(2.5nm) and Pt(6nm)$|$Co(2.5nm).}\label{fig4}
\end{figure}

In FM/NM bilayers we must additionally consider the fact that the transverse thermoelectric signal, $R^{H, \nabla T}_{2\omega}$, is mixed with $R^{H, SOT}_{2\omega}$. These two effects, however, can be separated in a quantitative way as shown in Ref.~\onlinecite{AvciPRB2014b} and explained briefly below. During an $xy$ scan of the magnetization, the FL-SOT gives a contribution to $R^{H}_{2\omega}$ proportional to $\cos 3\varphi + \cos \varphi$, whereas the AD-SOT and the ANE both give a contribution proportional to $\cos \varphi$. The AD-SOT and ANE contributions can be further separated by
considering that the AD-SOT induces dynamical oscillations of the magnetization, the amplitude of which is proportional to the magnetic susceptibility of the FM layer. The resulting $R^{H, SOT}_{2\omega}$ signal therefore depends on the susceptibility of the magnetization, which decreases with increasing external field, whereas the ANE contribution is constant provided that the magnetization is saturated along the field direction, as is the case in our measurements. The $\cos \varphi$ component of $R^{H}_{2\omega}$ is in fact a linear function of the inverse of the effective magnetic fields acting on the magnetization, $(B_{ext}+B_{DEM}+B_{PMA})^{-1}$, with slope proportional to the AD-SOT and intercept proportional to the ANE~\cite{AvciPRB2014b}. Figure~\ref{fig4} shows a plot of this component for two of the samples used in this study, Ta(6nm)$|$Co(2.5nm) and Pt(6nm)$|$Co(2.5nm). The intercept of the linear fit gives a non-negligible ANE-induced signal in the case of Ta ($R^{H, \nabla T}_{2\omega} = 1.1$~m$\Omega$), whereas a negligible ANE is found for the Pt sample ($R^{H, \nabla T}_{2\omega} = 0.02$~m$\Omega$). The corresponding MR of thermal origin is then given by $R^{\nabla T}_{2\omega} = \frac{l}{w} R^{H, \nabla T}_{2\omega}$. The above study is repeated for all the samples used in this work to separate the thermoelectric contributions from the measured second harmonic MR and reveal the pure USMR signal, given by $R^{USMR}_{2\omega}=R_{2\omega}-R^{\nabla T}_{2\omega}$.

\section{Absence of the USMR in single layer samples}
\label{Absent}

\begin{figure}[b]
  \centering
  \includegraphics[width=13 cm]{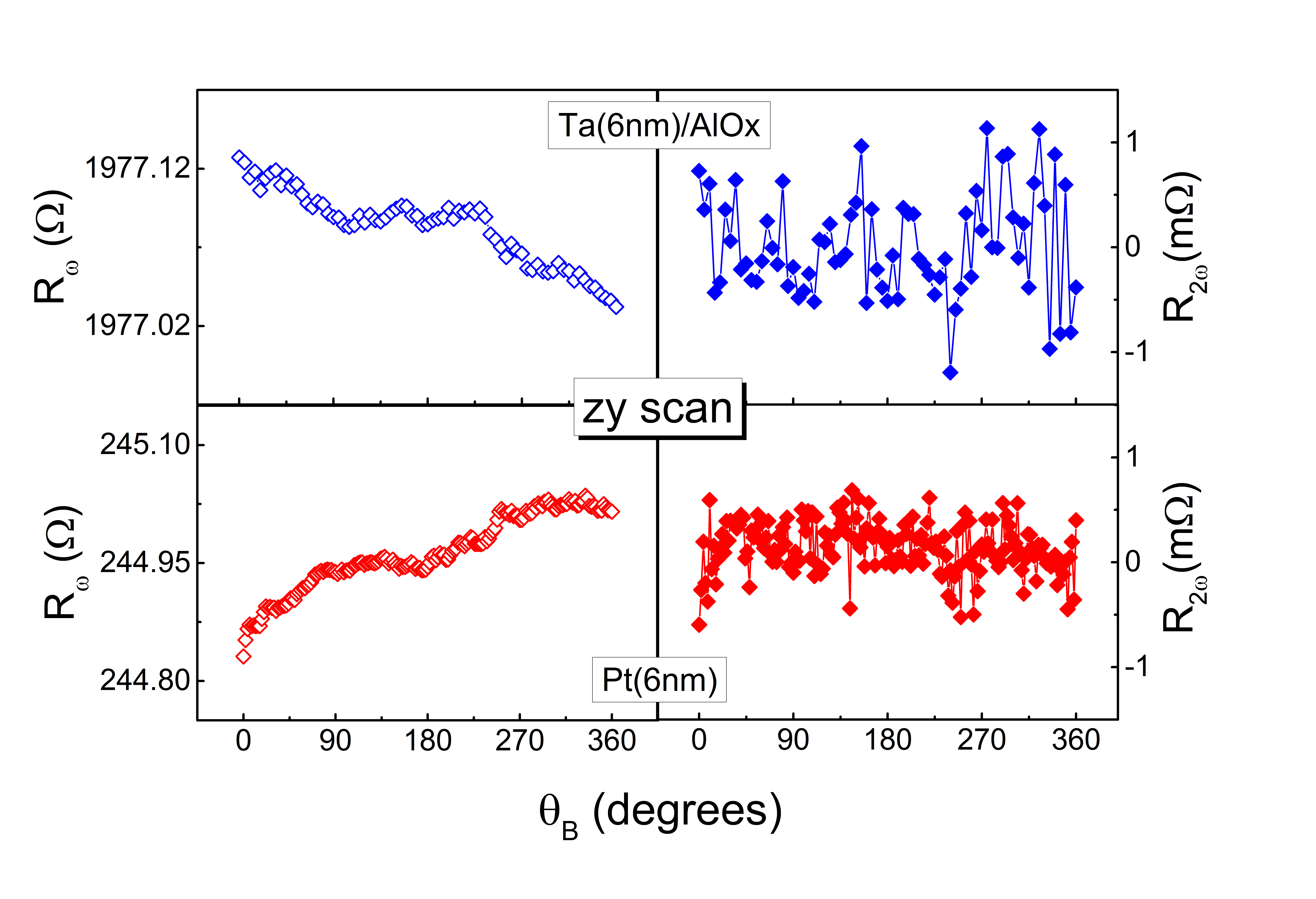}\\
  \caption{\textbf{Measurements of $R_{\omega}$ or $R_{2\omega}$ in single Ta and Pt layers.} First (left) and second (right) harmonic longitudinal resistance signals in Ta(6nm) (top) and Pt(6nm) (bottom) reference layers. Due to the high resistivity of Ta, the Ta data are averaged over five $zy$ scans performed with $B_{ext}=1.7$~T and $j=0.2\cdot10^{7}$~A/cm$^2$. For Pt, a single $zy$ scan is shown with $j=10^{7}$~A/cm$^2$ and $B_{ext}=1.9$~T.}\label{fig5}
\end{figure}

The measurements presented in Fig.~\ref{fig3} exclude the existence of a net USMR signal for a single Co layer. In order to further verify if the USMR is due to the Co/NM interface or to another nonlinear mechanism giving rise, e.g., to a self-induced MR induced by the SHE, such as that due to edge spin accumulation predicted by Dyakonov~\cite{DyakonovPRL2007}, we have grown two Ta(6nm) and Pt(6nm) reference layers and measured $R_{\omega}$ and $R_{2\omega}$ for these samples. Figure~\ref{fig5} shows the $zy$ scans for Ta (upper panels) and Pt (lower panels). The measurements are performed using a current density $j=0.2\cdot10^{7}$~A/cm$^2$ for Ta and $j=10^{7}$~A/cm$^2$ for Pt. The lower current density used for the Ta layer is due to the input voltage limit of our acquisition system and the high resistivity of Ta. Within the sensitivity of our measurements, we do not observe any systematic magnetic field dependence of either $R_{\omega}$ or $R_{2\omega}$, which excludes any influence from the NM layer alone in the USMR.

\section{USMR in T\lowercase{a}$|$C\lowercase{u}$|$C\lowercase{o} layers}

In order to elucidate the possible role of the FM-induced magnetization in the NM, we have inserted a Cu spacer with thickness 2 and 4~nm between Ta and Co. The choice of Cu is motivated by its large spin diffusion length as well as by its low and short-range induced magnetization. Ta(6nm)$|$Co(2.5nm) was chosen as a test system because of the larger absolute amplitude of the USMR relative to Pt. The aim of this study is twofold. First, to see if the USMR is still present when the Ta and Co layers are physically separated. Second, to examine if there is any influence of the induced magnetization on the reported USMR signal, although, judging by the change of sign of the USMR observed in the Ta$|$Co and Pt$|$Co layers, this appears unlikely. If the induced magnetization plays a role, we do not expect to measure a USMR signal after the insertion of the Cu spacer layers. However, if the USMR is due to the current-induced spin accumulation at the FM/NM interface, we expect to measure $R^{USMR}_{2\omega}\neq 0$ but reduced in amplitude with respect to Ta$|$Co, mainly due to current shunting in the Cu layer.

\begin{figure}[h]
  \centering
  \includegraphics[width=8 cm]{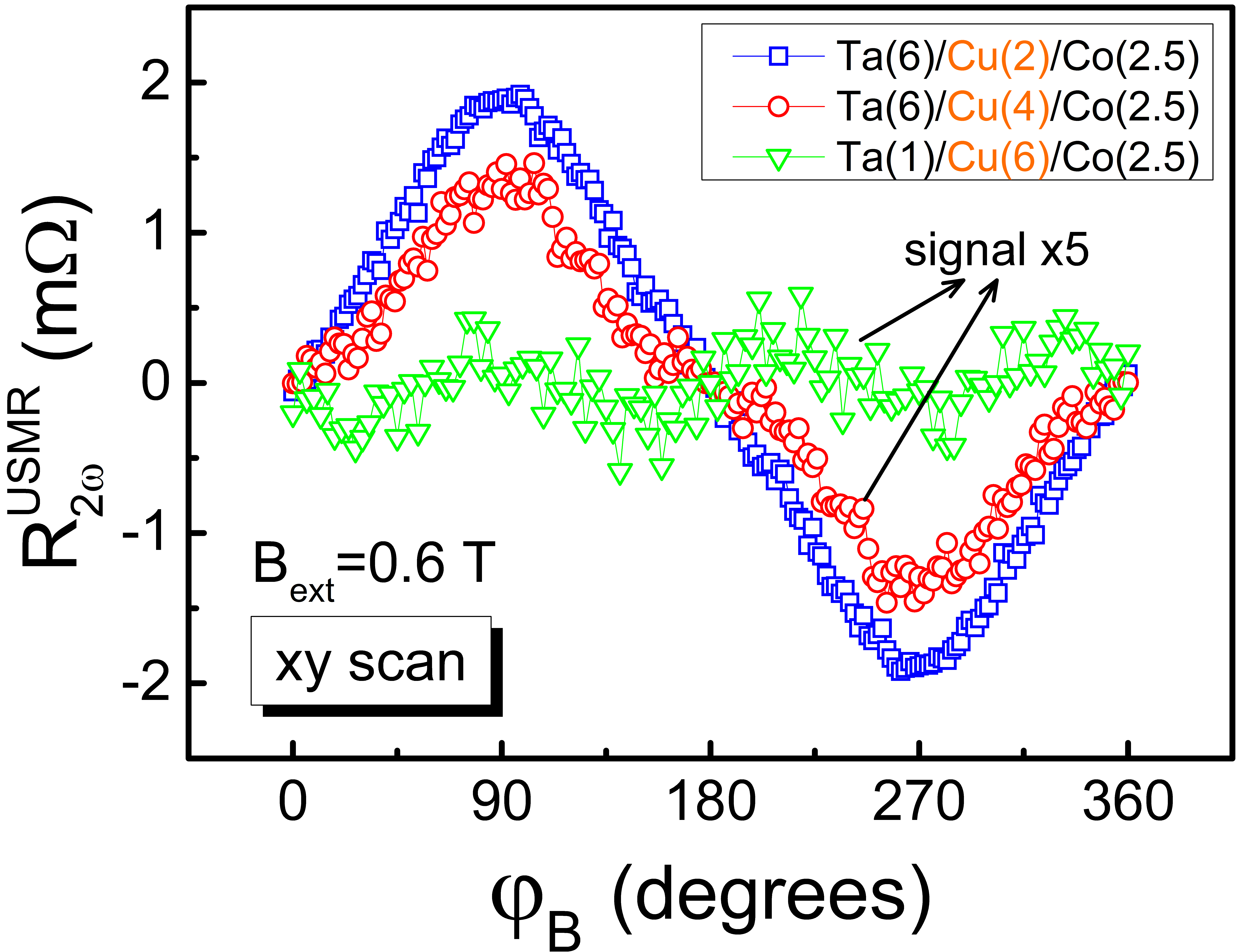}
  \caption{\textbf{Influence of a Cu spacer layer on the USMR signal of Ta$|$Co.} $xy$ scan of $R^{USMR}_{2\omega}$ measured in Ta(1,6nm)$|$Cu(2-6nm)$|$Co(2.5nm) trilayers with current density $j=10^7$~A/cm$^2$.}\label{fig6}
\end{figure}

Figure~\ref{fig6} shows the USMR of Ta(1,6nm)$|$Cu(2-6nm)$|$Co(2.5nm) layers. We observe a clear signal in Ta(6nm)$|$
Cu(2,4nm)$|$Co(2.5nm) that has the same sign as that measured in Ta(6nm)$|$Co(2.5nm). As expected, the USMR decreases significantly with increasing Cu layer thickness, from $1.9$~m$\Omega$ for Ta(6nm)$|$Cu(2nm)$|$Co(2.5nm) to $0.28$~m$\Omega$ for Ta(6nm)$|$Cu(4nm)$|$Co(2.5nm), and so does the resistivity of the devices, from $54.6~\mu\Omega$cm to $34.1~\mu\Omega$cm, showing that the shunting effect of Cu quickly dominates the conduction. The sample with Ta(1nm) has a resistivity of $16.9~\mu\Omega$cm and serves as a reference to show that the USMR vanishes if the conduction is dominated by the Co/Cu bilayer, due to the negligible SHE of Cu. We note that the 1~nm of Ta was deposited as an adhesion layer for Cu on the SiO substrate and that its conductivity is expected to be negligibly small due to oxidation and the much larger resistivity of Ta relative to Cu. Overall, these data confirm that the USMR is dominated by the spin accumulation at the FM/NM interface induced by the SHE in the heavy metal layer and that the induced magnetization in the NM does not play a significant role.

\section{Spin accumulation and potential shift induced by the SHE in a FM/NM bilayer}
There are essentially two effects that are important to interpret the USMR and model its dependence on the thickness of the NM layer. One is the buildup of the SHE-induced spin accumulation at the FM/NM interface, which occurs over a length-scale comparable to the spin diffusion length of the NM, and the other is a dilution effect due to reduced number of electrons that scatter at the interface relative to the total number of electrons participating to the conduction. We consider these two effects separately.

Our first point is the assumption that the USMR is proportional to the nonequilibrium spin accumulation at the FM/NM interface induced by the in-plane charge current. As explained in the text, this is motivated by the modulation of the conductivity mismatch that exists at the interface between a FM and a NM due to the SHE-induced change of majority and minority spin populations. In order to calculate the spin accumulation, we adopt the drift-diffusion approach~\cite{vanSon1987PRL,Valet1993PRB}. The spin accumulation is defined as $\boldsymbol{\mu}_s = \boldsymbol{\mu}^\uparrow - \boldsymbol{\mu}^\downarrow$, where $\boldsymbol{\mu}^\uparrow$ and $\boldsymbol{\mu}^\downarrow$  represent the spin-dependent electrochemical potentials for majority ($\uparrow$) and minority ($\downarrow$) electrons. We recall that the nonequilibrium magnetization in the NM is $\delta \mathbf{m} = -\mathcal{N}(\varepsilon_F)\mu_B \boldsymbol{\mu}_s$, where $\mathcal{N}(\varepsilon_F)$ is the density of states at the Fermi level and the minus sign stems from the opposite orientation of the magnetic spin moment and spin angular moment.

We consider a bilayer consisting of a NM with electrical conductivity $\sigma_N$, spin diffusion length $\lambda_N$, and thickness $t_N$, and a FM defined by the analogous quantities $\sigma_F$, $\lambda_F$, and $t_F$. The conductivity of the FM is assumed to be the sum of the independent conductivities for the majority and minority electrons: $\sigma_F = \sigma^\uparrow + \sigma^\downarrow$. Accordingly, the current flowing in the FM has a net spin polarization $P = (\sigma^\uparrow - \sigma^\downarrow)/(\sigma^\uparrow + \sigma^\downarrow)$. We define $x$ as the current direction and $z$ as the direction normal to the interface; the interface plane is situated at $z=0$. The source term for $\boldsymbol{\mu}_s$ is the spin current generated inside the NM by the SHE and propagating along $z$:
\begin{equation}
\mathbf{j}^{0}_{s}=j^{0}_{SH}(\hat{\mathbf{j}} \times \hat{\mathbf{z}})=-j^{0}_{SH}\hat{\mathbf{y}}= -\theta_{SH}\sigma_{N}E_x\hat{\mathbf{y}}\, ,
\label{js}
\end{equation}
 where $\hat{\mathbf{y}}$ represents the spin polarization direction, $E_x$ the electric field driving the charge current $\mathbf{j}$ through the bilayer, and $\theta_{SH}$ the spin Hall angle of the NM. The spin Hall angle of the FM is assumed to be zero. This approximation neglects the spin and charge accumulation induced by the anomalous Hall effect at the boundaries of the FM, which can be included in more detailed calculations but are not essential to the arguments developed here. Furthermore, we assume that the magnetization of the FM is saturated parallel to $\pm \hat{\mathbf{y}}$. In this case, there are no sink terms for the spin current inside the FM other than spin-flip relaxation; in particular, the spin current associated to the SOT via the real and imaginary part of the spin-mixing conductance $(G_{\uparrow\downarrow})$ vanishes\cite{Chen2013PRB}:
\begin{equation}
\mathbf{j}^{SOT}_s \sim \mathrm{Re}\{G_{\uparrow\downarrow}\}\boldsymbol{\tau}_{AD} + \mathrm{Im}\{G_{\uparrow\downarrow}\}\boldsymbol{\tau}_{FL} \sim \mathrm{Re}\{G_{\uparrow\downarrow}\}\mathbf{M}\times (\hat{\mathbf{y}} \times \mathbf{M}) + \mathrm{Im}\{G_{\uparrow\downarrow}\}(\mathbf{M} \times \hat{\mathbf{y}})=0 \, .
\label{jsSOT}
\end{equation}
With this assumption the spin diffusion equation governing the spin accumulation reduces to a one-dimensional problem:
 \begin{equation}
 \nabla^2\mu_{s_{F,N}}(z)=\frac{\mu_{s_{F,N}}(z)}{\lambda_{F,N}^2}\, ,
 \label{SpinDiffEq}
 \end{equation}
where $\mu_{s_{F,N}}$ is the $y$ component of the spin accumulation vector in either the FM or the NM layer. The spin current $\mathbf{j}_s = j_s\hat{\mathbf{y}}$ is then given by
 \begin{align}
 j_{s_{N}}(z) &=-\frac{\sigma_N}{2e}\partial_z(\mu_N^\uparrow - \mu_N^\downarrow) - j^{0}_{SH}  & \text{in the NM layer }\quad (z<0) \, , &\\
 j_{s_{F}}(z) &=-\frac{1}{e}\partial_z(\sigma^\uparrow\mu_F^\uparrow - \sigma^\downarrow\mu_F^\downarrow)  & \text{in the FM layer }\quad (z>0) \, . &
 \label{jsz}
 \end{align}
The general solutions of Eq.~\ref{SpinDiffEq} read
 \begin{align}
\begin{rcases}
\mu_N^\uparrow(z) &= A_N + B_Nz + C_N\exp(z/\lambda_N)+D_N\exp(-z/\lambda_N) \,&\\
\mu_N^\downarrow(z) &= A'_N + B'_Nz + C'_N\exp(z/\lambda_N)+D'_N\exp(-z/\lambda_N) \,&
\label{muN}
\end{rcases}
z < 0 \\
\begin{rcases}
\mu_F^\uparrow(z) &= A_F + B_Fz + C_F\exp(z/\lambda_F)+D_F\exp(-z/\lambda_F) \,&\\
\mu_F^\downarrow(z) &= A'_F + B'_Fz + C'_F\exp(z/\lambda_F)+D'_F\exp(-z/\lambda_F) \,&
\label{muF}
\end{rcases}
z > 0
\end{align}
The coefficients appearing in Eqs.~\ref{muN}, \ref{muF} are determined by imposing charge conservation $\nabla \cdot\mathbf{j} = 0$ in the two layers:
\begin{align}
& \partial_z^2 (\frac{\sigma_N}{2}\mu_N^\uparrow +  \frac{\sigma_N}{2}\mu_N^\downarrow) = 0  & z<0 \, , &\\
& \partial_z^2 (\sigma^\uparrow\mu_F^\uparrow +  \sigma^\downarrow\mu_F^\downarrow) =0  &  z>0 \, , &
 \label{jsz}
\end{align}
and by the following boundary conditions:
\begin{enumerate}
  \item The net charge flow perpendicular to the FM/NM interface is zero:
  \begin{equation}
  j_F^\uparrow + j_F^\downarrow = j_N^\uparrow + j_N^\downarrow = 0 .
  \label{B1}
  \end{equation}

  \item The spin-dependent current components are continuous across the FM/NM interface:
  \begin{equation}
  j_F^{\uparrow,\downarrow}(0)=j_N^{\uparrow,\downarrow}(0).
  \label{B2}
  \end{equation}

  \item The discontinuity of the electrochemical potentials of majority and minority electrons at the FM/NM interface is proportional to the spin-dependent boundary resistances $r^\uparrow$ and $r^\downarrow$, respectively:
      \begin{equation}
      \mu_F^\uparrow(0) - \mu_N^\uparrow(0) = er^\uparrow j^{\uparrow}(0) \quad \text{and} \quad \mu_F^\downarrow(0) - \mu_N^\downarrow(0) = er^\downarrow j^{\downarrow}(0).
      \label{B3}
      \end{equation}

  \item The spin current vanishes at the bottom of the NM layer and at the top of the FM layer:
      \begin{equation}
      j_{s_N}(-t_N)=j_{s_F}(t_F) = 0.
      \label{B4}
      \end{equation}
\end{enumerate}
In Eqs.~\ref{B1}-\ref{B3} above we have introduced the majority and minority current components, $j_F^{\uparrow,\downarrow} = \frac{\sigma_F^{\uparrow,\downarrow}}{e}\partial_z\mu_F^{\uparrow,\downarrow}$ and $j_N^{\uparrow,\downarrow} = \frac{\sigma_N}{2e}\partial_z\mu_N^{\uparrow,\downarrow} \mp \frac{1}{2}j^{0}_{SH}$, where the "majority" direction in the NM is defined with respect to that of the FM. The boundary condition~\ref{B2} implies the continuity of the spin current, $j_{s_F}(0)=j_{s_N}(0)$, which is  justified only in the absence of spin-flip scattering at the interface\cite{Valet1993PRB}. To simplify our model, we assume this to be the case even if spin-flip scattering at the Co$|$Pt interface can reduce the spin current by as much as a factor two\cite{bass2014jmmm}. In Eq.~\ref{B3} we retain the effects of interface scattering due to the different band structure of the NM and FM and diffuse scattering by
the roughness or chemical disorder of the interface, which are relevant in the theory of CIP-GMR\cite{hood94prb}. The spin-dependent boundary resistances are defined in analogy with the spin-dependent resistivity in bulk FM:
\begin{equation}
      r_{\uparrow(\downarrow)}=2r_b(1-(+)\gamma),
      \label{rb}
\end{equation}
where $r_b = (r_\uparrow+r_\downarrow)/4$ is the interface resistance (in units of $\Omega$m$^2$) and $\gamma = (r_\downarrow -r_\uparrow)/(r_\uparrow+r_\downarrow)$ the interfacial spin asymmetry coefficient\cite{Valet1993PRB}.
\begin{figure}
  \centering
  \includegraphics[width=14 cm]{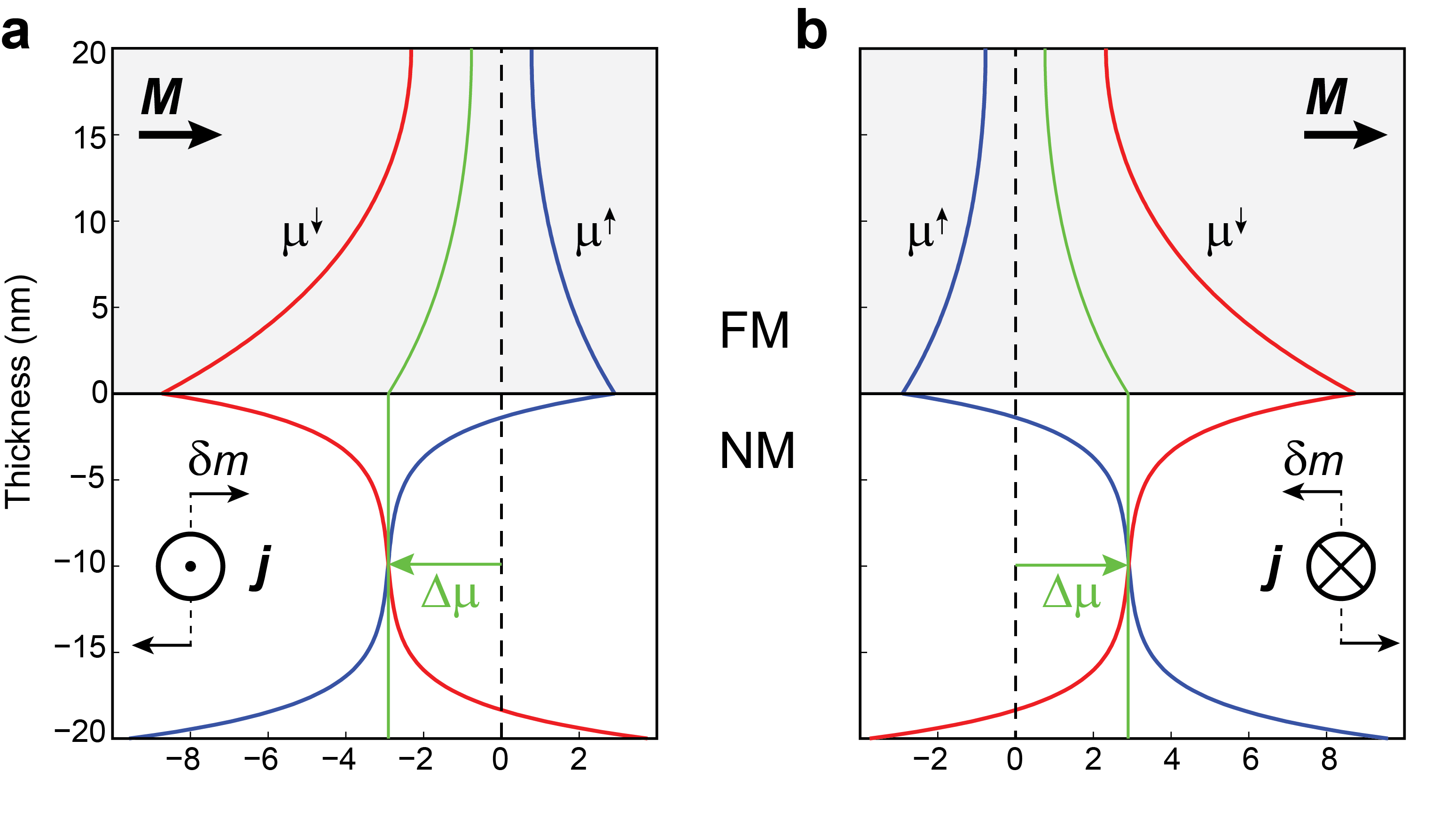}\\
  \caption{\textbf{Spin accumulation and modulation of the spin-dependent electrochemical potential by the SHE.} Profile of the electrochemical potential of majority ($\mu^\uparrow$, blue lines) and minority ($\mu^\downarrow$, red lines) electrons in proximity of the FM/NM interface calculated using Eqs.~\ref{muN}-\ref{muF} for \textbf{a,} positive and \textbf{b,} negative current. The electrochemical potential of the NM shifts relative to that of the FM as indicated by the green arrow. The direction of the magnetization is $\mathbf{M} \parallel \hat{\mathbf{y}}$ in both panels.}\label{fig7}
\end{figure}

Finally, we obtain
\begin{equation}
      \mu_{s_N}(0)=\mu_{s_N}^0\tanh\frac{t_N}{2\lambda_N}\frac{1+\frac{r_b}{\rho_F\lambda_F}(1-P^2)\tanh\frac{t_F}{\lambda_F}}
      {1+\left(\frac{\rho_N\lambda_N}{\rho_F\lambda_F}\coth\frac{t_n}{\lambda_N} - \frac{r_b}{\rho_F\lambda_F}\right)(1-P^2)\tanh\frac{t_F}{\lambda_F}},
      \label{muN0}
\end{equation}
where $\mu_{s_N}^0 = 2\,e \,\rho_N\lambda_N \,\theta_{SH}\,j$ is the bare spin accumulation due to the SHE that would be obtained for a single, infinitely thick NM layer and we have substituted the conductivity by the corresponding resistivity parameters $\rho_N = 1/\sigma_N$ and $\rho_F = 1/\sigma_F$. Note that in the limit $\rho_F \rightarrow\infty$ we recover the spin accumulation calculated for an insulating FM/NM interface\cite{Chen2013PRB,NakayamaPRL2013}. Differently from the latter case, however, the SHE induces a shift of the electrochemical potential of the NM layer, $\mu_{N} = (\mu_{N}^\uparrow + \mu_{N}^\downarrow)/2$, relative to that of the FM, $\mu_{F} = (\mu_{F}^\uparrow + \mu_{F}^\downarrow)/2$. Taking $\mu_F(\infty)=0$ as the reference level, we have
\begin{equation}
      \Delta\mu_N = \mu_N-\mu_F(\infty) =-(P+\gamma \tilde{r})\mu_{s_N}^0\tanh\frac{t_N}{2\lambda_N}\frac{1}{1+\tilde{r}}\frac{1}
      {1+\frac{\frac{\rho_N\lambda_N}{\rho_F\lambda_F}(1-P^2)\tanh\frac{t_F}{\lambda_F}\coth\frac{t_n}{\lambda_N}}{1-\tilde{r}}},
      \label{DeltamuN}
\end{equation}
where $\tilde{r}=\frac{r_b}{\rho_F\lambda_F}(1-P^2)\tanh\frac{t_F}{\lambda_F}$.

By measuring the resistivity of Co, Pt, and Ta reference layers and using published values for $P$, $\lambda_N$, $\lambda_F$, $r_b$, and $\gamma$, it is possible to estimate the relative weight of the different terms appearing in Eqs.~\ref{muN0} and~\ref{DeltamuN}. For the resistivity, we obtain $\rho_{Co} = 36.1$~$\mu\Omega$cm, $\rho_{Pt} = 33.4$~$\mu\Omega$cm, and $\rho_{Ta} = 228$~$\mu\Omega$cm. We take $P=0.31$ and $\lambda_{Co} = 30$~nm from CPP-GMR measurements of Co$|$Cu films with resistivity similar to ours\cite{piraux1998epjb}, $r_b = 0.74$~f$\Omega$m$^2$ and $\gamma = 0.53$ as recently measured for Co$|$Pt (Ref.~\onlinecite{bass2014jmmm}). Lacking similar measurements for Co$|$Ta, we assume the same interface resistance values as for Co$|$Pt. Further, we take $\lambda_{Pt} \approx  \lambda_{Ta} \approx 1.5$~nm (Refs.~\onlinecite{HahnPRB2013,ZhangAPL2013}). With these parameters and $t_F = 2.5$~nm, we calculate the products $\frac{\rho_N\lambda_N}{\rho_F\lambda_F}(1-P^2)\tanh\frac{t_F}{\lambda_F}= 0.0034$ (0.023) for Co$|$Pt (Co$|$Ta) and $\tilde{r}=0.005$. Inserting these products into Eqs.~\ref{muN0} and~\ref{DeltamuN}, we observe that, for $t_N\gtrsim \lambda_N$, the spin accumulation and the electrochemical potential shift of Co$|$Pt (Co$|$Ta) are approximated to within 1\% (3\%) by
\begin{equation}
      \mu_{s_N}(0) \approx \mu_{s_N}^0\tanh\frac{t_N}{2\lambda_N},
      \label{muN0approx}
\end{equation}
and
\begin{equation}
      \Delta\mu_N \approx -P\mu_{s_N}^0\tanh\frac{t_N}{2\lambda_N} .
      \label{DeltamuNapprox}
\end{equation}
More generally, we notice that, as long as $t_N\gtrsim \lambda_N$, $t_F < \lambda_F$, $\rho_N\lambda_N < \rho_F\lambda_F$, and $r_b < \rho_F\lambda_F$, both $\mu_{s_N}(0)$ and $\Delta\mu_N$ are mainly determined by the properties of the NM. Figure~\ref{fig7} illustrates the behavior of $\mu^\uparrow_{N,F}$, $\mu^\downarrow_{N,F}$, and $\Delta\mu_N$ as a function of $z$ calculated using Eqs.~\ref{muN}-\ref{muF} and the following parameters:
$j=10^7$~A/cm$^2$, $\theta_{\scriptscriptstyle{SH}}=0.1$, $P=0.5$, $\rho_N = \rho_F = 30$~$\mu\Omega$cm, $\lambda_N = 1.5$~nm, $\lambda_F = 10$~nm, $r_b = \gamma =0$. Figure~5 of the main text reports a similar calculation where $r_b = 0.74$~f$\Omega$m$^2$ and $\gamma = 0.53$ as appropriate for Co$|$Pt (Ref.~\onlinecite{bass2014jmmm}), which introduces a discontinuity between $\mu^{\uparrow,\downarrow}_{N}$ and $\mu^{\uparrow,\downarrow}_{F}$ at $z=0$.

\section{Current shunting through the NM and FM layers}

As the USMR is a pure interface effect, we expect a decrease of $\Delta R^{USMR}$ when the thickness of the NM and FM layers becomes larger than the respective spin diffusion lengths and the current is shunted away from the interface. This effect is similar to the "dilution" of the CIP-GMR observed in FM/NM/FM multilayers\cite{dieny92jpcm}. However, differently from CIP-GMR, the relevant lengthscale beyond which the dilution effect becomes significant here is the spin diffusion length rather than the electron mean free path.

We use a simple parallel resistor model to describe the current flow in a FM/NM bilayer. In this model, the "inactive" regions of the NM and FM layers are described by the resistances $R_N$ and $R_F$ whereas the "active" interface region is described by a linear, current-independent resistance $R_I$ in series with a nonlinear, current-dependent USMR resistance $r_s$. The sign of $r_s$ is determined by the cross product $\mathbf{j} \times \mathbf{M}$. Retaining only first order terms proportional to $r_s$, the equivalent resistance of $R_N$, $R_F$, and $R_I \pm r_s$ connected in parallel is
\begin{equation}
      R^{\pm} =  \frac{R_I}{1+R_I\frac{R_N+R_F}{R_N R_F}}\pm r_s \, ,
      \label{Rpar}
\end{equation}
which gives
\begin{equation}
      \Delta R^{USMR} = R^{+}-R^{-} =\frac{2r_s}{\left( 1+R_I\frac{R_N+R_F}{R_N R_F}\right)^2}.
      \label{DeltaR}
\end{equation}
To analyze the USMR dependence on the thickness of the NM layer alone, it is practical to include the constant resistance terms $R_F$ and $R_I$ in a single parameter $R_{FI}= R_F R_I / (R_F + R_I)$. We thus obtain
\begin{equation}
      \Delta R^{USMR} = \frac{2r_s}{\left( 1+\frac{R_{FI}}{\rho_N}\frac{w}{l}t_N\right)^2},
      \label{DeltaR2}
\end{equation}
and
\begin{equation}
      \frac{\Delta R^{USMR}}{R} = \frac{2r_s}{R_{FI}}\frac{1}{1+\frac{R_{FI}}{\rho_N}\frac{w}{l}t_N},
      \label{DeltaRR}
\end{equation}
where
\begin{equation}
R = \frac{R^+ + R^-}{2} = \frac{R_N R_{FI}}{R_N + R_{FI}} = \frac{R_{FI}}{1+\frac{R_{FI}}{\rho_N}\frac{w}{l}t_N},
\end{equation}
and $w/l$ is the ratio between the width and the length of the current line.

\section{Dependence of the USMR on the thickness of the NM layer}

As argumented in the main text and above, we expect the USMR to be proportional to $\Delta \mu$ times a dilution factor accounting for the shunting of the charge current by the NM layer. Combining Eqs.~\ref{muN0approx}, \ref{DeltaR2}, and \ref{DeltaRR} we obtain the following phenomenological expressions:
\begin{equation}
      \Delta R^{USMR} = A \frac{\tanh\frac{t_N}{2\lambda_N}}{\left( 1+\frac{R_{FI}}{\rho_N}\frac{w}{l}t_N\right)^2}, \quad
      \frac{\Delta R^{USMR}}{R} = \frac{A}{R_{FI}} \frac{\tanh\frac{t_N}{2\lambda_N}}{ 1+\frac{R_{FI}}{\rho_N}\frac{w}{l}t_N},
      \label{DeltaUSMR}
\end{equation}
where $A$ is a fitting parameter proportional to $P\rho_N \lambda_N \theta_{SH} j$ that accounts also for quantitative differences entering into the calculation of the resistance that escape our model. The values of the free parameters of the fits presented in Fig.~4 of the main text are $A=7$~m$\Omega$ (-14~m$\Omega$) for Ta$|$Co (Pt$|$Co), $R_{FI}/\rho_N=0.08$~nm$^{-1}$ (0.44~nm$^{-1}$), $\lambda_{Ta}= 1.4$~nm and $\lambda_{Pt}= 1.1$~nm.

%